\documentclass[aps,prl,reprint,superscriptaddress,nofootinbib,longbibliography]{revtex4-2}

\usepackage[T1]{fontenc}
\usepackage{graphicx}
\usepackage{amsmath,amssymb}
\usepackage{color}
\usepackage{hyperref}
\hypersetup{colorlinks,citecolor=blue,urlcolor=blue,linkcolor=blue}

\def\beq{\begin{equation}\begin{aligned}}
\def\eeq{\end{aligned}\end{equation}}

\begin{document}

\title{An All-Sky Search for Moving Gamma-Ray Sources with Fermi-LAT}

\author{Dou Lai}
\affiliation{PRIMES, Massachusetts Institute of Technology, Massachusetts 02139, USA}
\affiliation{Walter Payton College Preparatory High School, Chicago, IL, USA}

\author{James Unwin}
\affiliation{Department of Physics, University of Illinois Chicago, IL 60607, USA}
\affiliation{Rudolf Peierls Centre for Theoretical Physics, University of Oxford, Oxford OX1 3PU, UK}

\begin{abstract}

Nearby compact gamma-ray sources can move appreciably during the Fermi-LAT mission and thus evade searches that assume a stationary sky.  Compact dark matter halos, dark matter spikes around primordial black holes (PBH), and evaporating PBH provide potential examples.  We perform a novel all-sky search for steady point sources moving at $0.25^\circ/{\rm yr}-0.8^\circ/{\rm yr}$ in ten years of 1-100 GeV Fermi-LAT data. For a transverse speed of $200~\mathrm{km\,s^{-1}}$, this angular range probes distances of roughly $3.0\times10^3~\mathrm{AU}$ to $9.7\times10^3~\mathrm{AU}$.  We find 577 candidate tracks, consistent with $533\pm27.0$ tracks in time-scrambled skies. We place a model-independent 90\% upper limit of 80.8 recovered signal tracks. At peak calibrated sensitivity, this corresponds to fewer than 123 steady, full-sky sources with an incident $E^{-2}$ photon spectrum. We give the calibrated detection efficiency of the search to steady point sources and use it to derive illustrative constraints on both evaporating PBHs and PBHs surrounded by annihilating dark matter halos.
\end{abstract}

\maketitle

 \vspace{-3mm} \section{Introduction} \vspace{-3mm}

Conventional $\gamma$-ray searches assume that sources have fixed celestial coordinates. This approximation is excellent for extragalactic sources, and most Galactic objects, but it need not hold for nearby sources.  A source that moves appreciably during a multi-year $\gamma$-ray survey deposits photons along a path rather than at a single point, leading to a track in the spatiotemporal data. Such moving sources are likely to be missed or discarded by traditional stationary point source analyses.

Moving source candidates include close ordinary compact remnants, such as an uncatalogued pulsar or isolated accreting stellar-mass black hole, although the required proximity (typically within $\mathcal{O}(10^4)$ AU), makes such objects exceedingly rare and likely detectable at other wavelengths. Better motivated candidates come from dark matter scenarios or other `beyond the Standard Model' phenomena. 
Dark matter is typically expected to annihilate in regions of high density, and since the resulting $\gamma$-ray luminosity scales as density squared, small dense clumps -ultracompact minihalos (UCMHs)- can be both highly luminous and locally abundant \cite{Bertone:2004pz,Scott:2009tu,Josan:2010vn,Aslanyan:2015hmi}.
Primordial black holes (PBH) \cite{Hawking:1971ei,Zeldovich:1967lct,Carr:1974nx,Scholtz:2019csj}, if they exist, offer another prime candidate. Low mass PBHs radiate directly via Hawking evaporation \cite{Hawking:1974rv}.
Observational bounds require PBH sto be a subcomponent of dark matter over much of the mass range above $10^{-10}M_\odot$ \cite{Carr:2026hot}. In mixed PBH-particle dark matter scenarios, dense halos can form around PBH and dark matter pair-annihilations can source $\gamma$-rays  \cite{Lacki:2010zf,Eroshenko:2016yve,Adamek:2019gns}.

Proper motion was proposed as a distinctive signature of $\gamma$-ray microhalos \cite{Koushiappas:2006qq,Geringer-Sameth:2010vcs}, but has not previously been applied to real data. The only related Fermi-LAT search examined proper motion for preselected catalogue candidates \cite{Fermi-LAT:2018pfs}. More recently, signatures from PBHs transiting the Solar System have been examined using AMS positrons and electromagnetic observatories \cite{Klipfel:2025bvh,Klipfel:2026jrx}. These studies reinforce the motivation for treating motion explicitly, but do not perform an all-sky photon-level search for $\gamma$-ray tracks, as undertaken here.

Specifically, we analyze Pass 8 SOURCE-class  \cite{Fermi-LAT:2009ihh,Bruel:2018lac} Fermi-LAT photons with energies between 1 and 100 GeV from 2016 March 9 to 2026 March 9, in ten annual time bins. We find 577 candidate tracks, consistent with a background of $533\pm27.0$ tracks, and use this result to place a model-independent limit on recovered signal tracks. While this initial search finds no significant excess, it is limited by a number of factors as we discuss in the concluding remarks. As an illustration, we utilize these bounds to place constraints on PBH populations. The main focus of this Letter is to present the search methodology and the accompanying public codebase \cite{code}.

 \vspace{-3mm} \section{Search design} \vspace{-3mm}

The perpendicular velocity of a moving source maps onto its angular speed on the sky
\beq
 \mu=\frac{v_\perp}{r}
 =2.4\,\frac{\mathrm{deg}}{\mathrm{yr}}
 \left(\frac{v_\perp}{200~\mathrm{km\,s^{-1}}}\right)
 \left(\frac{10^3~\mathrm{AU}}{r}\right),
\eeq
where $v_\perp$ is the transverse speed and $r$ is the distance.  Angular speed is a critical factor in developing a moving-source search strategy. 
At fixed perpendicular velocity, lower angular speeds correspond to more distant objects. 
Notably, if the angular speed is sufficiently high, parallax may become detectable and must be taken into account, whereas if it is too low, the source is effectively stationary over the observation period.
To proceed, we select the angular velocity interval
\beq
0.25~\mathrm{deg/yr}\leq\mu\leq0.8~\mathrm{deg/yr}.
\eeq 
Assuming $v_\perp$ to be of order the characteristic halo velocity $v_\perp\simeq 200~\mathrm{km\,s^{-1}}$ the interval implies a sensitivity to objects within a distance shell between $3.0\times10^3~\mathrm{AU}$ and $9.7\times10^3~\mathrm{AU}$ (i.e.~roughly the region of the Oort cloud). The upper limit of the detectability window is also restricted by inverse-square reduction in flux, $F_\gamma=L_\gamma/(4\pi r^2)$, which implies that more distant sources are dimmer, and eventually too faint to detect.

\begin{figure}[t]
  \includegraphics[width=1\linewidth]{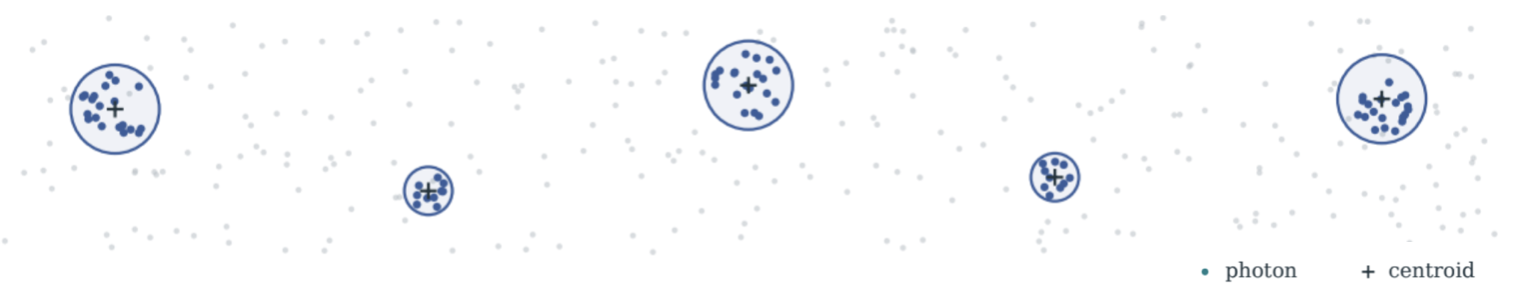}
\vspace{1mm}
  \includegraphics[width=1\linewidth]{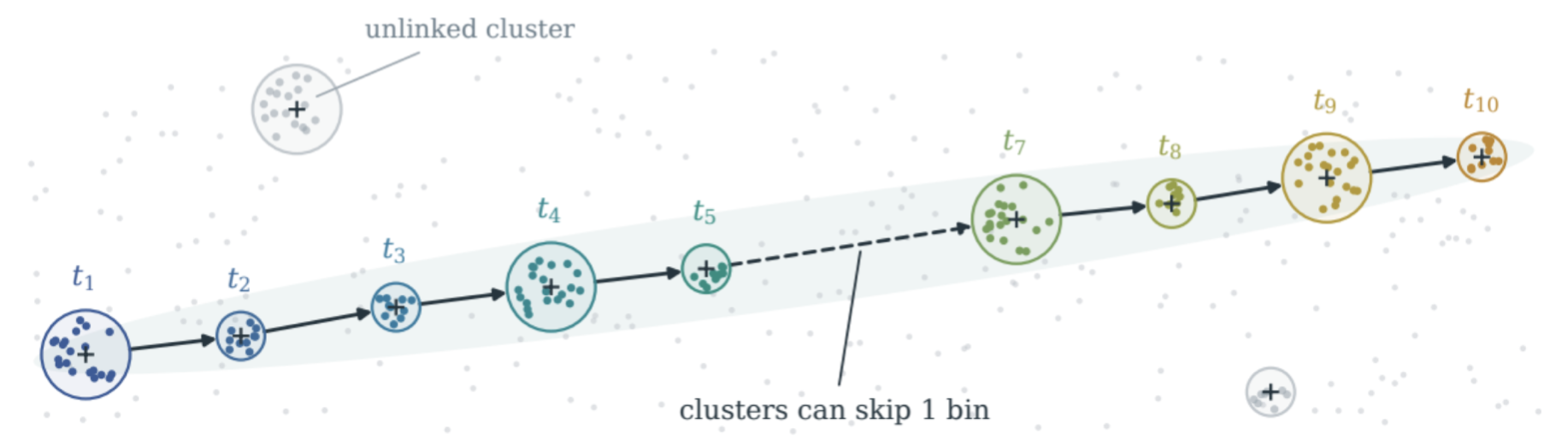}
\vspace{-4mm}
\caption{Schematic of the search. {\em Top.}~Within one annual time bin, photons ($\cdot$) are clustered on an energy-dependent angular scale; clusters that significantly exceed the local background are retained with their centroids ($+$). {\em Bottom.}~Centroids from successive bins $t_1,\dots,t_{10}$ are linked into an approximately linear trajectory; a link may skip one intervening bin (dashed), and a track must contain at least five clusters. Clusters not consistent with the trajectory are left unlinked (grey).}
\label{fig:search}
\vspace{-4mm}
\end{figure}

The sky is partitioned in 448 overlapping $18^\circ$-radius caps, the union of which covers the full sky. Each candidate track is assigned to a unique Voronoi cell
containing its midpoint.  For a given year and energy band, we apply a clustering algorithm to identify photon overdensities on an energy-dependent angular scale approximating the LAT point-spread function. The number of photons required to identify a possible source is adjusted according to the local background. Each candidate is then tested against the background, after masking known catalog sources, using the Li-Ma on/off significance \cite{Li:1983fv}. 
Surviving annual clusters are linked into ordered, nearly linear trajectories containing clusters in at least five annual bins.  Fig.~\ref{fig:search} shows a schematic of the methodology.

We estimate how often unrelated photons can mimic a moving source by repeating the full analysis on randomized versions of the sky. All selection criteria and statistical procedures were established using simulated sources and time-scrambled skies before examining the candidates found in the real data.
We mask regions containing known stationary sources from the FL16Y catalog \cite{Fermi-LAT:2022byn,Ballet:2026kpq}  and account for these excluded regions consistently when estimating the background.
The Supplemental Material  \cite{SM} provides technical details on the methodology.
To estimate the background, we create 60 randomized versions of the data, which we refer to as null campaigns. In each campaign, every photon is assigned a new arrival time drawn from the observed distribution in the same patch of the sky, while its measured direction, energy, and event classification remain unchanged. 
  
  To measure the sensitivity of the search, we separately add 2,250 simulated moving sources, one at a time, to the real photon data and determine how often they are recovered. These sources span a range of brightnesses, speeds, and sky locations. Recovery depends strongly on Galactic latitude; at the brightest calibrated setting it reaches 96\% at high latitude but only 32\% within $10^\circ$ of the plane. The efficiency entering the limits is therefore averaged over three latitude bands, weighted by their sky fractions \cite{SM}.
We measure how efficiently the search detects steady, point sources with an incident reference spectrum $dN/dE\propto E^{-2}$ across 1-100 GeV. The calibration folds this spectrum through the LAT's direction- and energy-dependent exposure when assigning detected photon energies, neglecting energy dispersion. This benchmark includes photons throughout the energy range. Sources with substantially different spectra may be detected with a different efficiency.
   A test using 120 simulated sources recovered 50, consistent with the 53.4 recoveries expected from the calibration \cite{SM}. 
  
\begin{figure}[t]
\centerline{
  \includegraphics[width=0.95\linewidth]{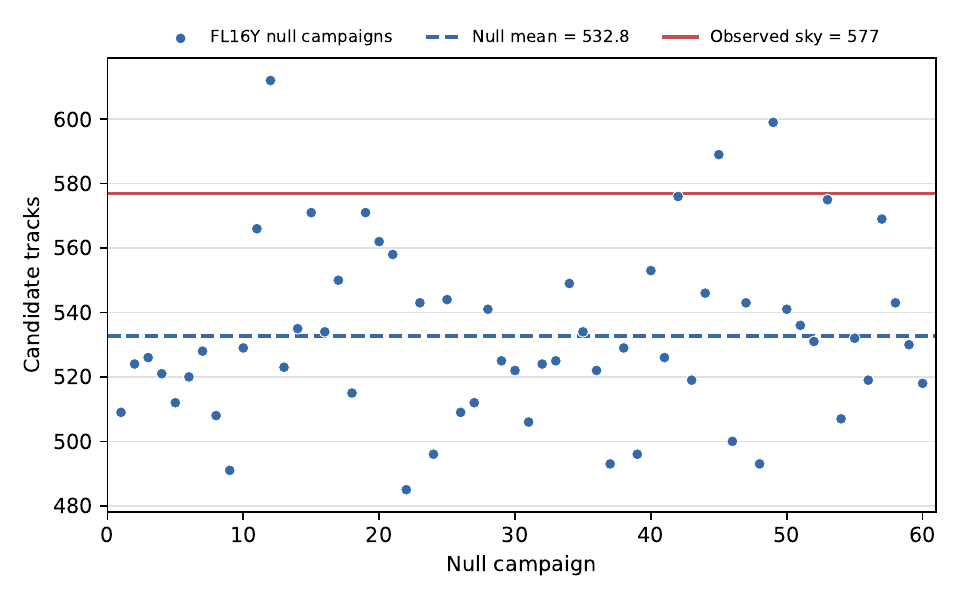}}
\vspace{-4mm}
\caption{Track counts over the 60 FL16Y null campaigns. The horizontal lines mark the observed FL16Y track count of 577 and the null mean of 532.8.}
\label{fig:null}
\vspace{-3mm}
\end{figure}

 \vspace{-3mm} \section{Model-independent result} \vspace{-3mm}

 Applying the search \cite{code} to ten years of 1-100 GeV Fermi-LAT Pass 8 SOURCE-class data, collected between 2016 March 9 and 2026 March 9  \cite{Fermi-LAT:2009ihh,Bruel:2018lac}, we identify 577 candidate tracks. Across 60 null campaigns, chance alignments of unrelated photons produce a
  mean of 533 tracks, with a campaign-to-campaign standard deviation of 27.0, see Fig.~\ref{fig:null}. Three of the 60 null campaigns produce at least as many tracks as the real data.  Accordingly, the plus-one probability implies a $p$ value
$ p=(3+1)/(60+1)\approx0.066$. This marginal $\sim1.5\sigma$ excess is not statistically significant.

We next determine how many tracks from genuine moving sources could be hidden within the observed count. We combine the measured background distribution with a hypothetical signal that contributes a fluctuating number of recovered tracks with mean $s$. We increase $s$ until the probability of obtaining 577 tracks or fewer falls to 10\%. This gives the 90\% upper limit $s_{90}=80.8$ recovered signal tracks. Thus, the present analysis cannot exclude a population contributing fewer than approximately 81 recovered tracks. 
 To convert this result into a limit on the underlying source population, we rescale $s_{90}$ by the recovery efficiency of genuine sources  $\epsilon$, defining $N_{90}:= s_{90}/\epsilon$.   At the most favourable tested setting, the search recovers approximately 82.3\% of simulated sources \cite{SM}. 
 
 These simulations contain 35 detected photons in each annual bin and move at $0.5~\mathrm{deg/yr}$. With this recovery rate the 80.8-track limit implies an upper bound of 98.2 sources over the full sky. 
  In addition, with only 60 null campaigns, the smallest probability that can be measured directly is  $1/(60+1)\approx0.016$. Calibrating a stronger discovery claim requires additional scrambled-sky campaigns.

\begin{figure}[t]
\centerline{
  \includegraphics[width=1.0\linewidth]{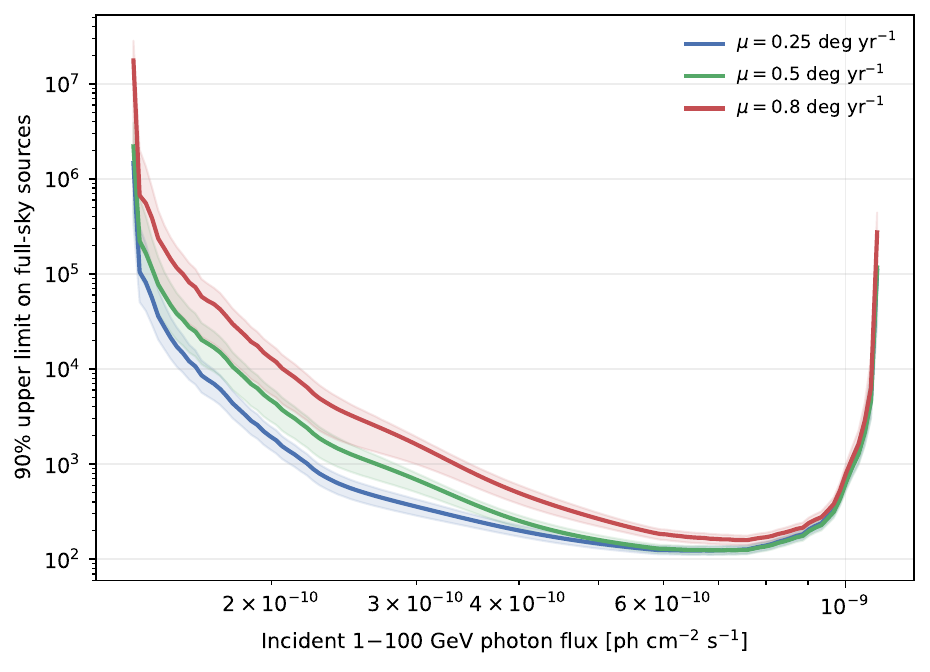}}
\vspace{-2mm}
\caption{Upper bound on the number of steady moving sources as a function of their average 1-100 GeV flux at 90\%CL, assuming different angular speeds.
The calibration assumes an incident $E^{-2}$ spectrum folded through the LAT exposure, as described in the text. The weakening at either edge marks the boundary of the calibrated response, not a fundamental decline in detectability.}
\vspace{-2mm}
\label{fig:result}
\end{figure}

Figure~\ref{fig:result} shows the corresponding 90\% upper limit on the number of moving $\gamma$-ray sources as a function of their common mean flux, for apparent angular speeds of ($0.25,~0.5,~0.8$) deg/yr. Each curve is conditional on every source having the indicated angular speed and should therefore be interpreted as a slice through this reference detection efficiency,  rather than as the kinematics of a realistic population. Physical populations will generally span a range of angular speeds because their members have different distances and differently projected transverse velocities. Limits of physical populations, therefore, must average the detection efficiency over the appropriate distance and velocity distributions, the PBH examples below perform this integration explicitly. After accounting for fluctuations in photon counts and the position-dependent LAT exposure, the strongest common-flux limit is 123 sources at $6.5\times10^{-10}~\mathrm{photons\,cm^{-2}\,s^{-1}}$ and $0.25~\mathrm{deg/yr}$, as can be seen in Figure~\ref{fig:result}.
  
   \vspace{-3mm} \section{Limits on PBH} \vspace{-3mm}

The model-independent result above can be recast as limits on specific populations of astrophysical objects. We consider two illustrative examples of annihilations within a dark matter halo around a PBH and a light PBH undergoing Hawking evaporation.
For a monochromatic population of PBHs that contributes a fraction $f_{\rm PBH}$ to the local dark
matter density $\rho_{\rm DM}$, the expected total number within a volume $V_{\mathrm{eff}}$ is
\beq
 N_{\rm det}=\frac{f_{\rm PBH}\rho_{\rm DM}}{M_{\rm PBH}}\,V_{\mathrm{eff}}(L_\gamma).
\eeq
Here the effective volume $V_{\mathrm{eff}}(L_\gamma)$ is the space over which a source of photon luminosity $L_\gamma$ would be detected, weighted by the probability that the search recovers it
\beq
V_{\mathrm{eff}}(L_\gamma)=
  \int {\rm d}\Omega\int_0^\infty r^2{\rm d}r\,
  \left\langle
  P_{\rm det}(L_\gamma,r,\Omega,v_\perp)
  \right\rangle,
  \eeq
  where $P_{\rm det}$ is the calibrated probability that the search recovers a source with luminosity $L_\gamma$, including the direction-dependent exposure, photon-count fluctuations, Galactic latitude, and angular speed $\mu$. The brackets denote an average over the assumed heliocentric transverse-velocity distribution.
The released effective volume \cite{code} integrates the detection efficiency $P_{\rm det}$ over distance, sky,  exposure, and halo-velocity distribution. 
 Setting the expected number of detected objects equal to the track-count limit gives at 90\% CL 
\beq
f_{\rm PBH}(M_{\rm PBH})=\frac{s_{90}M_{\rm PBH}}{\rho_{\rm DM}V_{\rm eff}[L_\gamma(M_{\rm PBH})]} .
\eeq
     
\begin{figure}[t]
\centerline{
  \includegraphics[width=1.0\linewidth]{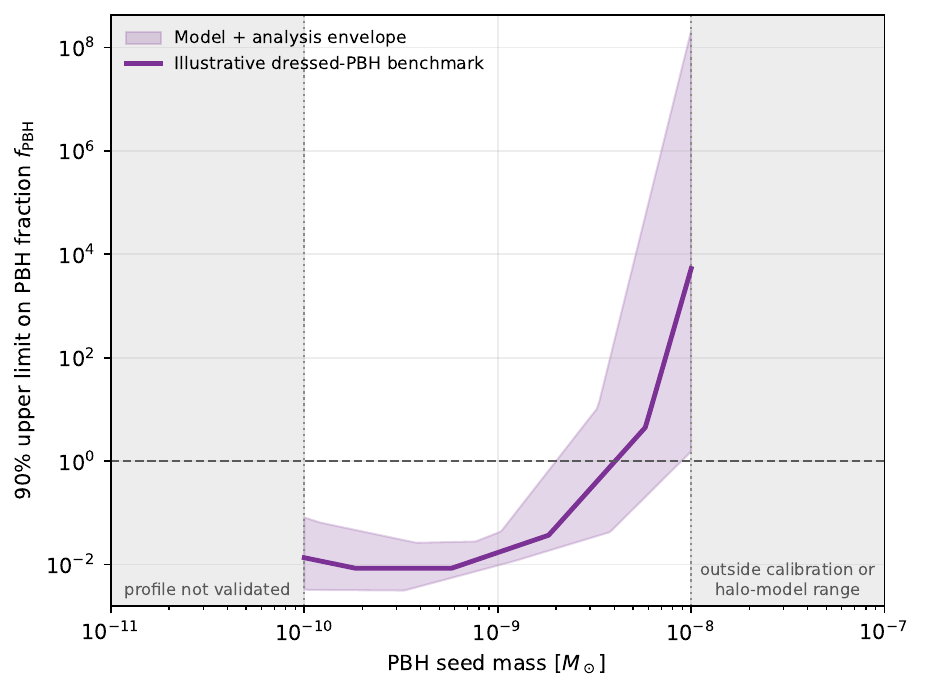}}
\vspace{-2mm}\caption{Upper limit on the fractional abundance of PBHs assuming dark matter annihilation within their associated halos. The solid curve shows our chosen example, while the surrounding band indicates variations in the source model and search sensitivity. Note, at masses higher than those shown, the sources fall outside the validated ranges of brightness and apparent motion. While for PBH masses lower than shown, the thermal kinetic energy of the dark matter particles can no longer be neglected, so the assumed $\rho\propto r^{-9/4}$ profile may cease to be reliable and a dedicated treatment is required \cite{Adamek:2019gns}.}
\vspace{-2mm}
\label{fig:pbh}
\end{figure}

Figure~\ref{fig:pbh}  shows the corresponding limits for a PBH dressed by a dark matter halo with a spike-profile  $\rho\propto r^{-9/4}$ \cite{Bertschinger:1985pd,Eroshenko:2016yve,Lacki:2010zf}, we take a 1 TeV particle annihilating to $b\bar b$ with $\langle\sigma v\rangle=3\times10^{-26}~\mathrm{cm^3\,s^{-1}}$.
The shaded band shows how the result changes under reasonable variations in the central halo density and the extent to which the halo has been stripped by tidal forces.
Across the calibrated interval $10^{-10}\leq M_{\rm PBH}/M_\odot\leq10^{-8}$, the strongest central bound is $f_{\rm PBH}<8.5\times10^{-3}$ at $M_{\rm PBH}\simeq2\times10^{-10}M_\odot$.  At this mass, the bound is comparable to microlensing constraints \cite{Niikura:2017zjd}. Diffuse $\gamma$-ray limit on this scenario provide stronger limits \cite{Lacki:2010zf,Chanda:2022hls,Gines:2022qzy,Chanda:2025bpl}, however, these analyses integrate over the cosmological population of dressed PBHs and thus depend on the formation history and redshift evolution, whereas the present limit is a direct constraint on the local population.

\begin{figure}[t]
\centerline{
  \includegraphics[width=1.02\linewidth]{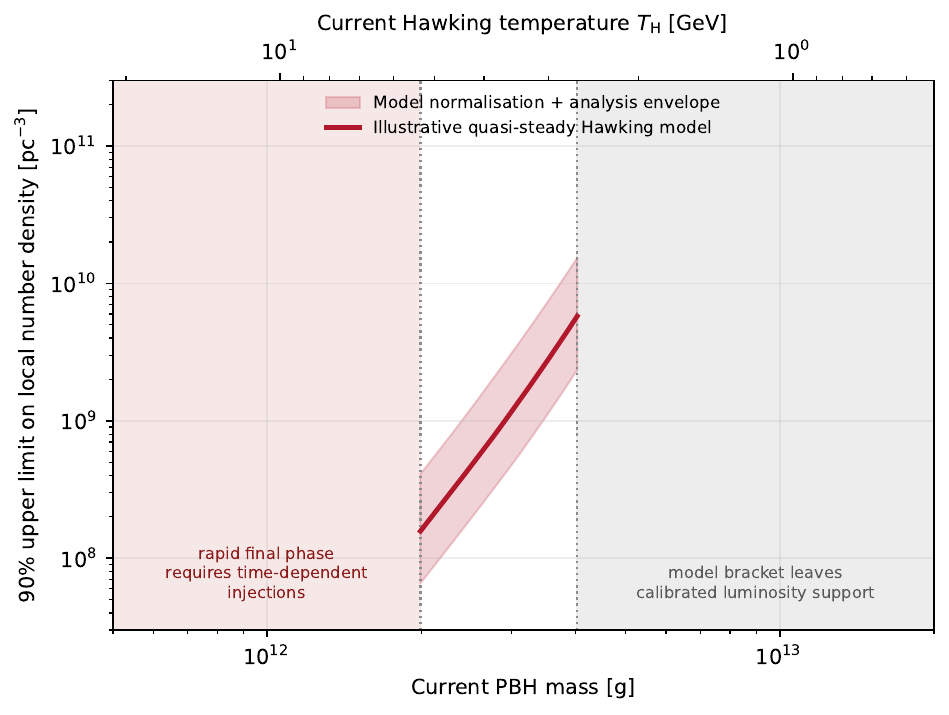}}
\vspace{-2mm}
\caption{Upper limit on the local number density of evaporating PBH as a function of mass.
The final rapid evaporation phase (left shading) requires time-dependent
injections. At higher masses (right shading), the predicted emission falls outside the brightness range tested by the search.}
\vspace{-2mm}
\label{fig:pbh2}
\end{figure}

Figure~\ref{fig:pbh2} applies the same search sensitivity to the $\gamma$-rays expected from Hawking evaporation, including photons produced by quark and gluon emission. We consider only the stage during which the emission changes slowly over the ten-year observation period. This corresponds to current PBH masses of $2-4\times10^{12}~\mathrm{g}$. In the standard evaporation scenario, this population is already more strongly constrained by the diffuse Galactic and extragalactic $\gamma$-ray backgrounds \cite{Carr:2026hot}, our current result is therefore illustrative rather than competitive. The final, rapidly brightening stage cannot be described by our steady-source calibration and would require a dedicated search for emission that changes with time. 

Both PBH curves are illustrative luminosity translations using the calibration for steady point sources with an incident $E^{-2}$ spectrum folded through the Fermi-LAT exposure. Their physical spectra determine the integrated 1-100 GeV luminosity, but were not used to generate spectrum-specific injection calibrations, fully spectrum-specific PBH exclusions would require dedicated injections. Despite these limitations the two PBH benchmarks indicate the validity and potential utility of this new astrophysical  $\gamma$-ray search implemented here.

   \vspace{-3mm} \section{Concluding Remarks} \vspace{-3mm}

  We have carried out the first all-sky photon-level search for persistent moving $\gamma$-ray sources in Fermi-LAT data. The 577 candidate tracks found in the data are consistent with the background measured from randomized skies, and therefore provide no evidence for a population of moving sources. The released code and effective-volume kernel \cite{code} allow these results to be applied to different source models without repeating the real search or null campaign, although different spectra or time dependence require new injection calibrations. The conversion of the model-independent track-count limit into a physical source-population limit is calibrated for steady, point-like sources in the 1-100 GeV band, using an incident $E^{-2}$ spectrum folded through the LAT exposure, as described above, and for angular speeds of $0.25$-$0.8\,\mathrm{deg\,yr^{-1}}$.
We demonstrated this recasting in the context of PBH populations, although the PBH studies presented here are intended as illustrative examples of the methodology, and we leave a full analysis to a dedicated paper.

  At a given transverse velocity, the angular-speed range   $0.25-0.8\,\mathrm{deg/yr}$ selects a finite range of distances. More distant objects appear stationary over the ten-year observation period, whereas sufficiently nearby objects move too rapidly and may follow curved, parallax-dominated paths (and are not constrained here). The present limits thus do not apply to either regime. Within the calibrated range, the same detection efficiency can be used to study other compact $\gamma$-ray populations, including ultracompact minihalos, by supplying the appropriate luminosity and motion model. 
 The Supplemental Material \cite{SM} contains full technical details.

By treating motion as a signature of the source rather than a complication, this work establishes a new way to search $\gamma$-ray data. Future work can develop this methodology by considering longer observing baselines in order to probe slower and more distant objects. Conversely, finer time bins with parallax-aware trajectories can open the search to nearby objects. Broader energy coverage, source-specific spectra, and larger null ensembles can extend the search beyond its present reach. The current null result is therefore a first constraint on this largely unexplored signal space, rather than its final implementation, and future searches may reveal populations that remain invisible in conventional stationary sky analyses.

\vspace{-4mm} \subsection*{Acknowledgments} \vspace{-5mm} 
J.U.~is supported by NSF Grant No. PHY-2209998 and thanks New College,
Queen's College, and the Rudolf Peierls Centre for hospitality, and Jakub
Scholtz for useful discussions. This work was completed as part of the MIT
PRIMES program. Claude and Codex assisted in aspects of the software development and manuscript preparation, the authors designed and validated the analysis and take responsibility for the work.

 \subsection*{Code and Data Availability} \vspace{-2mm} 

This study made use of public data and software from the Fermi Science
Support Center, as well as Astropy, NumPy, SciPy, scikit-learn, and Matplotlib.
LAT photon and spacecraft data and the FL16Y-v41 source list are publicly available from the Fermi Science Support Center. The analysis repository \cite{code} contains the frozen search and null products, effective-volume kernel, benchmark calculations, and scripts that reproduce the figures.



\onecolumngrid

\section{Supplemental Material}

This Supplemental Material gives the details needed to understand and reproduce the analysis summarized in the Letter and in the accompanying code \cite{code}. 
The text is organized as follows: Section~1 relates the flux and apparent motion of a nearby source to its luminosity, distance, and transverse velocity. Section~2 describes the LAT data selection, sky tiling, source masking, photon clustering, and track construction. Section~3 presents the randomized-sky background calibration, statistical inference, and potential backgrounds. Section~4 describes the artificial-source injections, recovery efficiency, Poisson count treatment, and validation tests. Section~5 constructs the effective-volume response, which is applied to the two illustrative PBH scenarios in Sections 6 and 7. 

\vspace{-4mm}
\section{1.~Flux, proper motion, and the detectable distance shell}
\vspace{-3mm}

We consider nearby gamma-ray emitters that appear point-like to Fermi-LAT. At fixed photon luminosity $L_\gamma$ and transverse speed $v_\perp$, the heliocentric distance $r$ determines both the observed photon flux $ F_\gamma(r)$ and the angular proper motion $\mu(r)$, given by
\beq 
F_\gamma(r)=\frac{L_\gamma}{4\pi r^2}, \hspace{1in}
\mu(r)= \frac{v_\perp}{r}.
\eeq
For a source which is well described in the constant angular velocity approximation, its position in a local tangent plane follows
\beq 
\theta(t)=\theta_0+\mu(t-t_0).
\eeq
As noted in the main body, for speeds of order halo velocity
\beq 
\mu\simeq 2.4\,\frac{\mathrm{deg}}{\mathrm{yr}} \left(\frac{v_\perp}{200\,\mathrm{km\,s^{-1}}}\right) \left(\frac{10^3\,\mathrm{AU}}{r}\right).
\label{eq:motion}
\eeq
The calibrated range $0.25\leq\mu\leq0.8\,\mathrm{deg\,yr^{-1}}$ implies a distance shell $3\times10^3~\mathrm{AU}\lesssim r\lesssim9.7\times10^3\,\mathrm{AU}$ if one assumes a characteristic speed $v_\perp=200\,\mathrm{km\,s^{-1}}$. These are not absolute distance limits, because a faster object at the same distance has a larger proper motion. The effective-volume calculation below therefore integrates over a velocity distribution rather than assigning every object one speed.
We approximate each source as moving across the sky at constant angular velocity and constant observed photon flux. For rectilinear space motion and constant intrinsic luminosity, $\dot\mu/\mu=\dot F_\gamma/F_\gamma=-2v_r/r$, where $v_r$ is the radial velocity, positive for recession. The importance of perspective effects is therefore controlled by $|v_r|T/r$, rather than by the angular speed alone. Annual binning may partially average the parallax displacement, but does not remove its contribution to the width of an annual photon concentration. The injection response and population limits below are evaluated in this straight-track, constant-flux approximation. Strongly nonlinear and Solar System bound trajectories are not calibrated.

The LAT point-spread function (PSF) describes the uncertainty in the reconstructed arrival direction of a detected photon. Even photons from a perfect point source appear spread around its true position; this spread is broader at low energies and narrower at high energies. Because a source moves during each one-year time bin, photons detected near the beginning and end of the year are centred at slightly different sky positions. 
The observed source image is therefore the PSF along a short path of length $\mu\Delta t$. It remains compact perpendicular to the motion but may be broadened along the direction of travel; for steady emission, the additional rms width along the path is $\mu\Delta t/\sqrt{12}$. The search consequently allows each annual photon concentration to be mildly elongated before linking its position across successive years. This effect is included directly in the simulated-source tests, which place photons along the source trajectory and then scatter each photon according to the energy-dependent PSF.

\vspace{-2mm}
\section{2.~LAT data and fixed analysis definition}

Fermi-LAT repeatedly surveys the full sky and records the reconstructed direction, energy, arrival time, and conversion type of each detected photon \cite{Fermi-LAT:2009ihh}. We use P8R3 SOURCE-class photons, including events converted in both the FRONT and BACK sections, together with the P8R3 SOURCE V3 description of the instrument's performance. The Pass 8 processing improves the rejection of charged-particle backgrounds that affected earlier data \cite{Bruel:2018lac}.

\vspace{-2mm}
\subsection{Event selection}
\vspace{-1mm}

We analyze photons recorded between mission elapsed times (MET) 479174404 and 794707205, corresponding to 2016 March 9 to 2026 March 9, and divide this ten-year interval into ten annual bins. We select 1-100 GeV P8R3 SOURCE-class events, including both FRONT and BACK conversion types (event-type mask 3), and use the P8R3 SOURCE V3 instrument response functions. We require a reconstructed zenith angle below $90^\circ$, use standard good-time intervals, and we remove events with missing or invalid position, energy, or arrival-time information. All data-selection choices were fixed before the final search was performed.
The lower energy threshold of 1 GeV in our chosen energy band avoids the very broad PSF at lower energies, where a compact source is difficult to distinguish from diffuse
emission. The upper threshold of 100 GeV defines the calibrated energy band. Standard good-time intervals remove periods when the data are unsuitable for science analysis, and a zenith angle below $90^\circ$ suppresses gamma rays from the bright Earth limb. We also reject events with nonfinite coordinates, energies, or times.

The full data set is too large to search in a single pass, so we divide the sky into $N_{\rm sky}=448$ overlapping processing caps. Each processing cap has a spherical radius of $18^\circ$, providing manageable data volumes together with substantial overlap and full-sky coverage. The cap centres form a quasi-uniform spherical Fibonacci tiling. They also define a non-overlapping set of nearest-centre spherical Voronoi ownership cells, which partition the sky and are used only to assign each completed track uniquely. We label the cap centres by $i=0,\ldots,N_{\rm sky}-1$. Their Galactic latitudes and longitudes, $b_i$ and $\ell_i$, are
\beq 
b_i=\arcsin \left[1-\frac{2(i+1/2)}{N_{\rm sky}}\right],\qquad\qquad \ell_i=i\pi(3-\sqrt{5}), 
\label{eq:fibonacci}
\eeq
where we use the golden angle $\pi(3-\sqrt{5})$ to distribute successive centres evenly in longitude.
The large overlaps between caps ensure that a source near the edge of a given sky cell still has all photons and its full background annulus available during clustering. The angular extent of each $18^\circ$ cap is 
\beq 
A_{18^\circ}=2\pi[1-\cos(18^\circ)]\approx 1010\,\mathrm{deg^2}. 
\label{eq:cap-area}
\eeq
There are $4.1\times10^4\,\mathrm{deg^2}$ in a sphere. We implement 448 caps, the union of which covers the entire sphere, so $f_{\mathrm{sky}}=1$.

Because the caps overlap, each photon lies within approximately 11 caps on average, and candidate tracks are initially constructed independently in each cap. The track search reduces hundreds of millions of recorded photon events to a manageable set of candidate moving tracks. It proceeds in three main steps. First, photons detected close together on the sky, within the same energy band and year, are grouped when they form an excess over the surrounding background. Next, groups from different years are linked when their changing positions are consistent with a coherent trajectory. These first two steps are illustrated in Fig.~1 of the main text. Finally, candidates that do not follow a sufficiently ordered path are rejected, and candidates satisfying the predeclared duplicate criteria are combined. By using information from several years, the search can identify a source that is too faint to stand out in any single year but repeatedly appears at positions tracing an ordered path across the sky.

 For each annual photon cluster, we estimate the local background using a ring of surrounding sky. The $18^\circ$ processing cap is sufficiently large that this ring is never truncated at the cap boundary for a track assigned to that cap. Any parts of the ring that overlap the masked regions around known catalogue sources are removed from both the background photon count and the sky area used to calculate the background rate.

Within a single cap, two reconstructed tracks are treated as duplicates when they share at least 80\% of the annual clusters in the shorter track. Only one representative of each such pair is retained; tracks below this threshold remain separate candidate hypotheses. Once track construction is complete, each surviving candidate is retained only in the cap whose centre is closest to the track's midpoint. As an additional safeguard against duplication caused by overlapping caps, tracks reconstructed in different caps are grouped when they share at least three time bins and their positions differ by less than $0.6^\circ$ in every shared bin. Only one representative of each group is retained. These are operational definitions of duplication: two tracks that share some detections but do not satisfy either criterion remain separate candidate hypotheses and need not represent distinct physical sources. Midpoint ownership must be assigned after track construction because it depends on the completed trajectory. The same procedure is applied to the real data, randomized skies, and source-injection tests.

 All search choices described below were established during development using simulated sources, randomized skies, and earlier catalogue products. Before examining the final FL16Y candidate list, we fixed the event selection, source masks, clustering and linking thresholds, ownership and duplicate rules, injection design, and statistical procedure. No threshold was adjusted after the number or properties of the 577 real-sky candidate tracks were known. 
 
\vspace{-2mm}
\subsection{Annual clustering and adaptive seeding}
\vspace{-2mm}

We identify annual concentrations of photons using an adaptive form of density-based spatial clustering (DBSCAN) \cite{Ester}. This method does not assume how many sources are present or where they are located. Instead, it identifies photons with enough nearby neighbours, joins connected groups of these photons, and then includes less crowded photons around the edges. This is well suited to an all-sky search in which the photons from a moving source may be spread slightly along its direction of travel during a one-year bin.
 Photon events are divided into five energy bands (in GeV):
\beq
\notag [1,~2.15], \qquad [2.15,~10], \quad [10,21.5], \qquad [21.5,46.4], \qquad [46.4,100].
\eeq 
The reconstructed direction of a LAT photon is less precise at low energy than at high energy. Photons from a low-energy point source can therefore appear spread over a larger area of sky, whereas using the same large search radius at high energy would include unnecessary background photons. Thus we assign each band an angular neighbourhood radius defined by 
\beq 
\varepsilon(E_*)= \max\left[0.8(E_*/\mathrm{GeV})^{-0.8},0.1\right]^\circ,
\label{eq:epsilon}
\eeq
where $E_*$ is the reference energy for that band. Two photons separated by less than $\varepsilon$ are treated as neighbors during clustering. We normally choose $E_*$ to be the geometric mean of the band boundaries. For the $2.15-10$ GeV band, however, we use $E_*=3.16\,\mathrm{GeV}$, giving the wider radius $\varepsilon\simeq0.32^\circ$ needed to accommodate its lower-energy photons. The same radius $\varepsilon$ defines the area around a cluster used to test for an excess. 

For each photon considered as the starting point of a cluster, we estimate the nearby background density using a surrounding ring that begins at least $2\varepsilon$ away and extends to $5\varepsilon$. We scale this density to the area of the $\varepsilon$ neighbourhood to obtain $\mu_{\rm loc}$, the expected number of unrelated background photons within the clustering radius.\footnote{An earlier version instead used the average density across the full processing cap. In a fixed-position diagnostic, it recovered 49 of 80 simulated sources in the lowest local-density quarter but none of 80 in the highest-density quarter.} Specifically, for each photon $p$ considered as the starting point of a cluster, we estimate the local background using photons from the same annual time bin and the same energy band. Let $N_{{\rm ann},p}$ be the number of such photons in the portion of the annulus extending from $2\varepsilon$ to $5\varepsilon$ around $p$. Let $A_{{\rm ann},p}$ be the area of this annulus and $A_{\varepsilon,p}$ the area within angular distance $\varepsilon$ of $p$. We define
\beq 
\rho_{{\rm loc},p}=\frac{N_{{\rm ann},p}}{A_{{\rm ann},p}}, \qquad \mu_{{\rm loc},p} =\rho_{{\rm loc},p}A_{\varepsilon,p} =N_{{\rm ann},p}\frac{A_{\varepsilon,p}}{A_{{\rm ann},p}} .
\eeq
Thus $\mu_{{\rm loc},p}$ is the estimated number of unrelated photons within the clustering neighbourhood of the test photon.

Keeping the background ring $2\varepsilon-5\varepsilon$ away from the possible source prevents its central concentration of photons from substantially increasing the threshold it must pass. If the unrelated background photons are approximately Poisson distributed, their typical count fluctuations are of order $\sqrt{\mu_{\rm loc}}$. Because the starting photon is itself included in the observed count, the background-only expectation is $1+\mu_{\rm loc}$. We therefore require 
\beq
n_{\rm core}\geq \max\left[4,\, \mu_{\rm loc}+2.5\sqrt{\mu_{\rm loc}}+1\right],
\label{eq:seed}
\eeq 
where $n_{\rm core}$ is the integer number of photons within $\varepsilon$, including the starting photon. The minimum of four prevents pairs or triplets from forming clusters in very sparse regions, while brighter backgrounds require a correspondingly larger concentration. Photons satisfying Eq.~(\ref{eq:seed}) form the main structure of a cluster, and additional nearby photons are attached around its edges.

This is only an initial cluster-forming condition, not a claim of $2.5\sigma$ detection significance. Completed clusters must subsequently pass the catalogue-masked on/off test described below, while the rate at which chance alignments survive the complete search is measured directly using randomized skies. 

\vspace{-2mm}
\subsection{Local significance and catalog masking} \vspace{-2mm}

For each annual photon cluster, we evaluate its significance using the on/off method of Li and Ma \cite{Li:1983fv}. We define the position of each cluster using the spherical photon mean. Each photon direction is represented by a unit vector $\mathbf n_k$, and the cluster direction is
\beq 
\widehat{\mathbf n}_{\rm cl} =\frac{\sum_{k=1}^{N}\mathbf n_k} {\left|\sum_{k=1}^{N}\mathbf n_k\right|}. 
\eeq
This direction defines the centre of the on and off regions. The on region is a disc of radius $\varepsilon$, while the off region is the surrounding annulus from $2\varepsilon$ to $5\varepsilon$. This nearby annulus estimates the local background while excluding the central concentration of photons associated with the candidate cluster. Let $N_{\rm on}$ and $N_{\rm off}$ denote the numbers of photons from the same annual time bin and the same energy band as the cluster in the usable on and off regions, respectively. Here $N_{\rm on}$ includes every selected photon in the on region, not only the photons assigned to the cluster by DBSCAN. The usable areas $A_{\rm on}$ and $A_{\rm off}$ include the same processing-cap boundary and catalogue-mask exclusions as the corresponding photon counts, and we define $\alpha=A_{\rm on}/A_{\rm off}$. For a positive excess, the Li-Ma significance is
\beq 
S_{\rm LM}=\Bigg\{2\Bigg[N_{\rm on}\ln \left( \frac{1+\alpha}{\alpha}
\frac{N_{\rm on}}{N_{\rm on}+N_{\rm off}} \right) +N_{\rm off}\ln \left((1+\alpha) \frac{N_{\rm off}}{N_{\rm on}+N_{\rm off}}\right)\Bigg]\Bigg\}^{1/2}. 
\label{eq:lima}
\eeq
We set $S_{\rm LM}=0$ when $N_{\rm on}\leq\alpha N_{\rm off}$, and only clusters with $S_{\rm LM}\geq2.5$ are retained for linking across successive years. This statistic ranks local excesses. Note that $S_{\rm LM}$ measures how unusual one annual photon concentration is relative to its immediate background. It is not a global significance because the search examines many overlapping sky positions, energy bands, years, and possible trajectories. We therefore apply the complete, unchanged search to randomized skies and compare their resulting track populations with the real data. This comparison accounts for the many possible false tracks and the correlations among them. 

We mask known stationary sources using the full FL16Y-v41 catalogue, without imposing a flux threshold \cite{Fermi-LAT:2022byn,Ballet:2026kpq}. A candidate cluster is rejected if it lies within $\max(0.5^\circ,\varepsilon)$ of a catalogue source, and a slightly larger region, $\max(0.5^\circ,2\varepsilon)$, is excluded when estimating the surrounding background. We remove both the photons and the corresponding sky area from this estimate; otherwise the inferred background density would be biased. A candidate is rejected if fewer than ten background photons or less than 25\% of the otherwise available annular area remains after catalogue masking. The same procedure is applied to the real data, randomized skies, and source-injection tests. Regions close to catalogue sources therefore act as permanent blind spots, whose loss of sensitivity is included in the measured injection-recovery efficiency. 
A physical source may produce surviving clusters in more than one energy band. Within each year and processing cap, we rank these clusters by $S_{\rm LM}$. After retaining the highest-ranked cluster, any lower-ranked cluster whose centroid lies within the larger of their two values of $\varepsilon$ is treated as a duplicate. Only the higher-ranked representative is retained, while the contributing energy bands are recorded. This procedure is repeated until every cluster has been considered, after which the surviving annual clusters are linked across time. \vspace{-4mm}

\subsection{Track linking and duplicate reduction} 
\vspace{-2mm}

To link annual clusters, we represent them as points in a graph ordered by time. Candidate $j$ can follow candidate $i$ if their bin-index difference is $g_{ij}\in\{1,2\}$ and their angular separation satisfies $d_{ij}\leq(1\,\mathrm{deg\,yr^{-1}})\Delta t_{ij}$. Thus one link may skip one intervening year. The algorithm first favours paths containing the largest number of annual detections and then compares their scores. Each possible partial track $T$ is assigned a score $Q(T)$ that rewards significant annual photon concentrations while penalizing large jumps and missing years. A track containing only cluster $i$ is initialized with $ Q(\{i\})=S_{{\rm LM},i}$. If cluster $j$ is appended to the end of $T$, we denote the extended track by $T\Vert j$ and calculate its new score as
\beq 
Q(T\Vert j)=Q(T)+S_{{\rm LM},j} -\left[\frac{d_{ij}}{v_{\rm link}\Delta t_{ij}} +0.5(g_{ij}-1) \right].
\eeq
Here $v_{\rm link}=1~\mathrm{deg}$ per 365-day analysis year is the maximum angular speed allowed when forming a link. The first penalty is therefore the inferred angular speed between the two clusters expressed as a fraction of this maximum. Thus, a locally stronger cluster increases the track score, whereas a larger angular step or a skipped year lowers it. The score is used to rank competing ways of linking the annual clusters, although its value should not be interpreted as a statistical significance. Valid candidate tracks are required to contain at least five detections and at most two skipped bins in total. The graph is searched using dynamic programming. For each possible endpoint and each accumulated number of skipped bins, partial tracks are ranked first by their number of annual detections and then by $Q$. We retain the four highest-ranked extended states for each endpoint and skip count, together with the single-cluster state that allows a new track to begin at that endpoint. Predecessor pointers are retained so that every surviving endpoint state can be reconstructed as a complete track.

We next test whether the annual cluster positions resemble steady motion. We first calculate the spherical mean direction $\widehat{\mathbf n}_{0}$ of the track positions and construct orthonormal east and north vectors, $\mathbf e_{\rm E}$ and $\mathbf e_{\rm N}$, in its tangent plane. Each cluster direction $\mathbf n_k$ is mapped to coordinates
\beq 
x_k=\frac{\mathbf n_k\cdot\mathbf e_{\rm E}}{\mathbf n_k\cdot\widehat{\mathbf n}_{0}}, 
\qquad 
y_k=\frac{\mathbf n_k\cdot\mathbf e_{\rm N}}{\mathbf n_k\cdot\widehat{\mathbf n}_{0}} .
\eeq
This unit-vector construction is insensitive to the $0^\circ/360^\circ$ right-ascension boundary and remains well defined near the celestial poles. We subtract the mean projected position and determine the direction along which the resulting tangent-plane positions are most extended. A moving source should form a narrow, nearly straight sequence, whereas random shifts around a stationary position are more likely to spread in multiple directions. If $\lambda_\parallel$ and $\lambda_\perp$ are the positional variances parallel and perpendicular to the preferred direction, we define
\beq
A_{\rm tube}=1-\frac{\lambda_\perp}{\lambda_\parallel}
\eeq
and require $A_{\rm tube}\geq0.7$. We also require the positions along this direction to correlate with time by at least 0.85, using the absolute Pearson correlation coefficient. The
first condition selects line-like tracks; the second requires the source to progress along the line in a consistent temporal order.\footnote{An earlier requirement trialed during development that successive steps have similar angular speeds was removed after injection tests showed that ordinary uncertainty in the measured cluster positions could reject genuine constant-speed sources. This quantity is still recorded for diagnostic purposes but is not used as a selection cut.} Surviving tracks are ranked first by the number of annual detections and then by their linking score $Q$. Subsequently, we remove tracks within the same processing cap that are constructed from substantially the same annual clusters. A track $T$ is treated as a duplicate of a higher-ranked track $T'$ when
\beq
\frac{|C(T)\cap C(T')|}{\min[|C(T)|,|C(T')|]}\geq0.8,
\eeq
where $C(T)$ is the set of annual clusters belonging to $T$. Thus two tracks are considered duplicates when they share at least 80\% of the clusters in the shorter track. We apply this
rule only after the motion tests so that an invalid extension of a genuine shorter track cannot remove the valid track.

\vspace{-3mm}\section{3.~Randomized-sky null and statistical inference}
\vspace{-2mm}

We next discuss the utilization of randomized versions of the observed sky to measure how often unrelated photons form apparent moving tracks and thereby determine the search background. We then compare the real candidate population with this empirical null distribution, assess its statistical significance, and derive an upper limit on the number of recovered signal tracks. 

\subsection{Null construction}

Random concentrations of diffuse photons can sometimes be linked into an apparently coherent path. Because the number of possible paths is too large for a simple analytic trials correction, we measure these false candidates by repeating the full search on 60 randomized skies. The null construction must remove the original association between direction and time while retaining the strong spatial nonuniformity of the gamma-ray sky, the spectrum and source environment of each region, and the local empirical distribution of photon arrival times.

Each photon is assigned a new arrival time sampled from the day-binned arrival-time distribution in the same local region of the sky, while preserving its identity and all its other measured properties. 
Specifically, for each null campaign, each photon is assigned to one of the 448 non-overlapping Voronoi regions. These regions have a mean area of $92~\mathrm{deg}^2$, corresponding to an effective radius of approximately $5.4^\circ$, and no boundary lies more than $7.4^\circ$ from its centre. For each photon, we retain its measured direction, energy, event class, and FRONT/BACK conversion type, and replace only its arrival time. The new time is drawn with replacement from the empirical daily time distribution of all quality-selected 1-100 GeV SOURCE-class photons in that region, a day is selected in proportion to its observed photon count and the time is then drawn uniformly within that day. 
The time distribution includes photons of all energies and both conversion types; the reassigned times are not matched by energy or conversion type. The randomization is defined reproducibly using each photon's RUN\_ID and EVENT\_ID, ensuring that copies of the same physical event in overlapping analysis caps receive the same randomized time.

Each campaign is then passed through the entire track search, ownership assignment, and duplicate reduction. The resulting campaign count measures how many tracks survive the complete search after reassigning photon arrival times while holding their other measured properties fixed. Sixty campaigns were completed.  The first 30 randomized skies were used to construct three possible descriptions of the background-count distribution. Their predictions were then tested against the independently reserved final 30 skies using predeclared calibration, coverage, and accuracy criteria. All 60 campaigns were subsequently combined to form the final empirical background distribution.

If $m$ of the $K$ randomized campaigns contain at least as many tracks as the real sky, we define the finite-campaign probability 
\beq 
p_+=\frac{m+1}{K+1}.
\eeq
The added one prevents a probability of zero when none of a finite number of randomized campaigns exceeds the observation.
We used the first 30 campaigns to compare three possible descriptions of the background. The campaign counts vary more than expected from an ordinary Poisson distribution, their variance is $1.37$ times their mean. Thus we use the measured distribution of randomized-sky counts directly, rather than imposing a narrower theoretical form.

\subsection{Count upper limit}

We set an upper limit by ascertaining how many genuine tracks could be hidden within the observed total. For every measured background count, we add a hypothetical signal whose number of recovered tracks fluctuates around a mean $s$. As $s$ increases, obtaining no more than the observed 577 tracks becomes less likely. Specifically, for a Poisson signal $S$ with mean $s$, we define
\beq 
F_s(N_{\rm obs})=\frac{1+\sum_{i=1}^{K} P[S\leq N_{\rm obs}-B_i]}{K+1}.
\eeq
This treats the observed count as $N_{\rm obs}=B+S$, with a Poisson signal count $S$ independent of the background count $B$.\footnote{Because the randomized skies are constructed from the observed photons, any photons from a genuine moving-source population would also enter the randomized campaigns. Reassigning their arrival times destroys the coherent trajectory but does not remove the photons, which could still contribute to chance clusters or tracks. The reported interval assumes that this contribution does not materially shift the randomized sky track count distribution. Any upward shift of the randomized-background distribution would tighten the upper limit at fixed observed count. The magnitude of any signal-induced bias have not been explicitly calibrated.}
The one-sided 90\% upper limit on the recovered signal $s_{90}$ is given by the nonnegative solution of $F_s(N_{\rm obs})=0.1$. A larger mean signal would give less than a 10\% chance of observing so few tracks.
We obtain an upper limit from the combined background distribution measured across all 60 randomized campaigns, i.e.~we do not calculate 60 separate limits and average them. We tested this procedure using 20,000 simulated repetitions for each of six possible signal strengths and four background variation levels, including the variation measured in the data. In all 24 tested combinations, the fraction of repetitions in which the upper limit covered the injected signal was consistent with the nominal 90\% coverage and satisfied the predeclared acceptance criterion, including for the background variation measured in the data.

\begin{table}[t]
\begin{ruledtabular}
\begin{tabular}{lcccc}
Candidate & $(\mathrm{RA},\mathrm{Dec})_{\rm start}$ & $(\mathrm{RA},\mathrm{Dec})_{\rm end}$
& $\langle\mu\rangle$ ($\mathrm{deg\,yr^{-1}}$) & $p_{\rm global}$ \\
\hline
ROI301-C001 & $(84.963,-7.427)$ & $(83.712,-5.494)$ & 0.255 & 0.180 \\
ROI301-C003 & $(84.963,-7.427)$ & $(86.981,-8.941)$ & 0.313 & 0.492 \\
ROI280-C001 & $(86.324,-1.071)$ & $(86.798,0.329)$ & 0.164 & 0.557 \\
ROI301-C002 & $(84.963,-7.427)$ & $(83.712,-5.494)$ & 0.255 & 0.672 \\
ROI211-C004 & $(308.791,39.592)$ & $(308.937,38.657)$ & 0.188 & 0.967 \\
ROI266-C001 & $(313.841,31.583)$ & $(314.296,31.268)$ & 0.056 & 0.984 \\
ROI232-C003 & $(312.197,37.260)$ & $(308.937,38.657)$ & 0.487 & 0.984 \\
ROI280-C002 & $(86.402,-1.815)$ & $(88.042,1.345)$ & 0.395 & 0.984 \\
ROI298-C001 & $(53.304,31.117)$ & $(54.501,31.147)$ & 0.114 & 0.984 \\
ROI280-C003 & $(86.324,-1.071)$ & $(85.586,-2.103)$ & 0.159 & 0.984 \\
ROI238-C001 & $(91.480,16.195)$ & $(91.303,14.718)$ & 0.165 & 0.984 \\
ROI293-C001 & $(80.290,8.033)$ & $(81.955,6.472)$ & 0.253 & 0.984 \\
ROI267-C001 & $(91.917,-7.116)$ & $(91.470,-5.715)$ & 0.184 & 1.000 \\
ROI285-C001 & $(63.640,28.308)$ & $(66.444,26.876)$ & 0.319 & 1.000 \\
ROI289-C001 & $(286.378,-37.224)$ & $(285.699,-37.281)$ & 0.068 & 1.000 \\
\end{tabular}
\end{ruledtabular}
\vspace{-2mm}
\caption{The 15 tracks with the largest summed Li-Ma scores. Reconstructed start and end positions are given as equatorial right ascension and declination in degrees. The global $p_{\rm global}$ is defined by Eq.~(\ref{eq:candidate-p}).
Tracks with identical start and end positions can have different intermediate annual detections. ROI301-C001 and ROI301-C002 are related track hypotheses that share six of eight annual clusters, including both endpoints, but follow different intermediate clusters. They are retained separately by the predeclared 80\% within-region duplicate criterion. \label{tab:candidates}}
\vspace{-2mm}
\end{table}

We also test whether the properties of the real candidate tracks differ from those found in the randomized skies. These comparisons consider track score, angular speed, duration, photon count, and the number of contributing energy bands. Each complete campaign is treated as one independent realization, avoiding the assumption that tracks within the same sky are independent. We compare the real-sky discrepancy with those obtained when each randomized campaign is treated as though it were the real sky, and correct the resulting probabilities for testing several properties. Finally, the randomized skies define a high-score follow-up threshold, both the number of candidates above this threshold and the score of the most extreme candidate are evaluated against the same campaign-level background.

After applying the ownership and duplicate removal rules, the real sky contains 577 candidate tracks. The 60 randomized campaigns contain an average of 533 tracks, with a campaign-to-campaign standard deviation of 27.0. The observed count lies $(577-533)/27.0\approx1.63$ campaign-to-campaign standard deviations above the randomized-sky mean. Three of the 60 null campaigns contain at least as many tracks as the
real data, giving the primary plus-one probability
\beq
p=\frac{3+1}{60+1}\approx 0.0656,
\eeq
which is equivalent to $1.51\sigma$ for a one-sided Gaussian tail. Thus this indicates a $1.63\sigma$ fluctuation above the null mean, while its directly calibrated significance is $1.5\sigma$; neither constitutes statistically significant evidence for a moving-source population. The distributions of track score, angular speed, duration, photon count, and contributing energy bands are also consistent with the randomized skies after accounting for the five comparisons.

Individual candidate scores are calibrated against the maximum score in each
complete null campaign, rather than against the much larger pooled set of
tracks. For this comparison, we use the summed Li-Ma score
\beq
{\cal S}(T)=\sum_{i\in C(T)}S_{{\rm LM},i}.
\eeq 
For a candidate with score ${\cal S}$
\beq 
p_{\rm global}({\cal S})= \frac{1+\sum_{k=1}^{60}I({\cal S}^{\rm max}_{\rm null,k}\geq{\cal S})}{61}. \label{eq:candidate-p}
\eeq
This comparison includes the full-sky search trials. Table~\ref{tab:candidates}
lists the 15 tracks with the largest ${\cal S}$. Even the highest-scoring candidate is not statistically significant, 10 of the 60 randomized skies contain a chance track with an equally high or higher score, giving the plus-one probability $p_{\rm global}=11/61\approx 0.18$. We therefore do not interpret any individual track as evidence for a moving source. The candidate list includes all tracks passing the search cuts and is not restricted to the angular-speed interval used for the injection calibration. The physical limits below use only the calibrated response over $0.25\leq\mu\leq0.8\,\mathrm{deg\,yr^{-1}}$.

\vspace{-4mm}
\subsection{Potential Backgrounds}
\label{sec:variability}
\vspace{-2mm}

The time scramble removes genuine stationary-source curves as well as proper motion. A variable stationary source can brighten in selected years, shift its measured centroid, and mimic a slow trajectory in real data without an equivalent object in the randomized-sky background. This is a limitation of the randomized-sky model, not of the injection response.
Before examining the results, we specified two possible signs of contamination by variable stationary sources. First, such sources might preferentially produce slow apparent tracks, so we counted tracks with angular speeds $0.25~\mathrm{deg~yr^{-1}}\leq\mu<0.35~\mathrm{deg~yr^{-1}}$. Second, we counted tracks lying just outside the normal catalogue veto, but within a further $0.5^\circ$, of an FL16Y source classified as variable. The real data contain 159 slow tracks, compared with an average of 133 in the randomized skies. Only two of the 60 randomized campaigns show an equally large or larger count, giving $p=3/61=0.049$; after accounting for the two tests, this becomes $p_{\rm adj}=0.0984$. By contrast, 87 real tracks lie near known variable sources, compared with a randomized-sky average of 94 ($p=0.8$), and the overall distribution of track-source separations is also consistent with the randomized skies ($p=0.16$). 
The data therefore provide no evidence that FL16Y-listed variable sources produce the candidates, while contamination from uncatalogued variable sources remains untested. We consequently neither claim that variability leakage is present nor that it has been excluded.

We also highlight that the four tracks with the smallest $p_{\rm global}$ in Table \ref{tab:candidates} lie in the direction of the Orion molecular cloud complex. The three ROI301 tracks lie towards Orion A, with the related tracks ROI301-C001 and ROI301-C002 ending near M42, while ROI280-C001 ends near M78 in Orion B. These clouds exhibit extended gamma-ray emission associated with cosmic-ray interactions with interstellar gas \cite{Ackermann:2012ik}. Such structured emission provides a plausible background environment in which chance photon concentrations may be linked into apparent tracks. This spatial coincidence was identified after the search and has not been assigned a statistical significance. The randomized skies retain the observed photon directions and hence this static spatial structure. We make no additional candidate rejection or change to the reported limits on this basis.

\vspace{-5mm}
\subsection{Catalog update}
\vspace{-2mm}

The search was initially developed and tested using the 4FGL-DR4 source catalogue. We fixed the event selection, clustering rules, track requirements, source-injection design, and statistical procedure before applying the search with the newer FL16Y catalogue. This separation prevented choices made during development from being adjusted to the final catalogue-masked candidate list.

\begin{figure}[t]
\centering
\includegraphics[width=0.55\columnwidth]{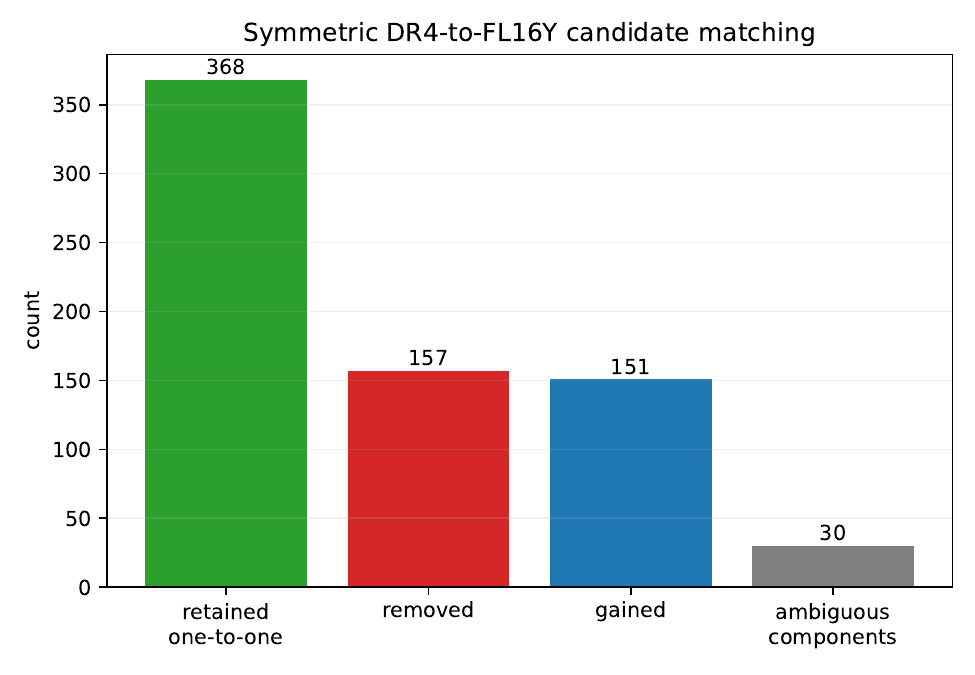}
\vspace{-3mm}
\caption{Candidate-level accounting for the DR4-to-FL16Y catalog update.
The current search uses only the FL16Y branch.}
\vspace{-2mm}
\label{fig:cat}
\end{figure}

 To isolate the effect of the catalogue update, the DR4 and FL16Y analyses use the same sky regions, search code, and randomized photon times; only the catalogue mask changes. The number of real-sky tracks after the duplicate removal decreases from 584 with DR4 to 577 with FL16Y, while the average randomized-sky count decreases from 535 to 533. Comparing the candidate lists, 368 tracks are retained, 157 DR4 tracks are removed, and 151 FL16Y tracks are new. The remaining 59 DR4 tracks and 58 FL16Y tracks form 30 groups that cannot be matched unambiguously. Individual tracks can change because the updated catalogue alters both which clusters are rejected near known sources and which photons are available for estimating the surrounding background.

Figure~\ref{fig:cat} shows how real and null candidates move among acceptance states under the catalog update. The resulting 90\% upper limit on recovered signal tracks changes from 87.0 to 80.8. All final detection efficiencies, physical limits, and abundance constraints reported in this work use FL16Y. Results obtained with DR4 are retained only to show the effect of the catalogue update and the robustness of the results over two catalogues.

\vspace{-3mm}
\section{4.~Validation}
\vspace{-3mm}

In this section, we explain how artificial moving sources are used to measure the efficiency with which the complete search recovers a genuine signal across different brightness levels, angular speeds, and sky environments. We then validate the resulting response, including the effects of the LAT PSF, Poisson fluctuations in detected photon counts, annual exposure variations, and catalog masking. These injected sources are processed using the same fixed FL16Y catalogue mask as the real and randomized data.\footnote{Because the FL16Y catalogue is held fixed, an injected source cannot create a new catalogue entry and subsequently veto itself. This possible self-veto was tested for the preceding detector-level-$E^{-2}$ calibration, no production injection reached the nominal catalogue threshold, and an intentionally conservative treatment in which possible threshold crossings were counted as complete losses changed the headline limits only slightly, except on the already calibration-limited high-luminosity turnover.}
 
\vspace{-3mm}
\subsection{Production design and PSF}
\vspace{-3mm}

We measure the search efficiency by adding simulated moving sources with known properties to representative regions of the real sky and processing them in exactly the same way as the data. Because the position of each simulated source is known in every year, we compare the reconstructed candidates directly with its expected trajectory. The source is counted as recovered when a candidate overlaps that trajectory in at least three annual bins and the median separation between the reconstructed and expected positions is no greater than $0.6^\circ$. Sources lost during annual clustering, catalogue masking, background testing, track linking, or duplicate removal are all counted as missed. The resulting efficiency therefore measures the performance of the complete search.

We use $R$ to label the apparent brightness of an injected source, defined as the exact number of artificial photons added to each of the ten annual bins before any search selections are applied. An injection with value $R$ therefore contains $10R$ photons over the complete observing interval. This fixed count is a controlled calibration quantity, not a physical flux or the number of photons remaining after the search cuts. We test five photon counts $ R=(8,12,18,25,35)$, being the exact number of artificial photons added to each annual bin before any search selections are applied. We also test three angular speeds, $ \mu=(0.25,0.5,0.8)\,\mathrm{deg\,yr^{-1}}$. The simulated position changes continuously, including during each annual bin. Fixing the photon count provides a controlled measurement of the search efficiency, although a real steady source will not produce exactly the same number of detected photons every year.

For the calibration, we assume an incident reference spectrum $dN/dE\propto E^{-2}$ between 1 and 100 GeV. The energies assigned to detected artificial photons are drawn from this spectrum after weighting by the position-dependent exposure
\beq
p_{\rm det}(E\mid\Omega)=
\frac{E^{-2}{\cal E}_{10}(E,\Omega)}
{\displaystyle\int_{1\,{\rm GeV}}^{100\,{\rm GeV}}
E'^{-2}{\cal E}_{10}(E',\Omega)\,{\rm d}E'} .
\eeq
Recall that ${\cal E}_{10}(E,\Omega)$ is defined in Eq.~(\ref{eq:E10}) and is evaluated in the direction of the simulated source. This expression accounts for the fact that the probability of detecting a photon is energy dependent. Each sampled energy is recorded as the energy of a given artificial photon, sets its clustering band, and selects the appropriate energy-dependent PSF.

To preserve realistic observing conditions, each artificial photon is created by copying the instrumental properties of a real LAT event. Each artificial photon is constructed from a real LAT event selected from the appropriate annual bin. Its arrival time, FRONT or BACK conversion type, and viewing geometry are retained, while its energy is redrawn and its direction is placed along the moving-source trajectory and displaced according to the corresponding P8R3 SOURCE V3 double-King PSF.\footnote{We validated the double-King injection sampler against the official LAT \texttt{gtobssim} simulator using 110 paired sources at $R=18,25,35$ and $\mu=0.5\,\mathrm{deg\,yr^{-1}}$ over a common eight-bin 2016-2023 interval. The paired recovery rates were statistically consistent. For catalogue-stable pairs, the changes in inferred inverse efficiency were 7.90\%, 8.15\%, and 3.87\%, all below the predeclared 10\% threshold for a material difference. We therefore apply no correction to the injection efficiency.} This distribution contains both a narrow central peak and a broader tail. Including the tail is important because photons reconstructed unusually far from the source can affect both annual clustering and track linking.\footnote{An earlier treatment employed a Gaussian approximation, but this was found to produce too few widely displaced photons.}

\begin{figure}[t]
\centering
\hspace*{-5mm}
\includegraphics[width=1.0\columnwidth]{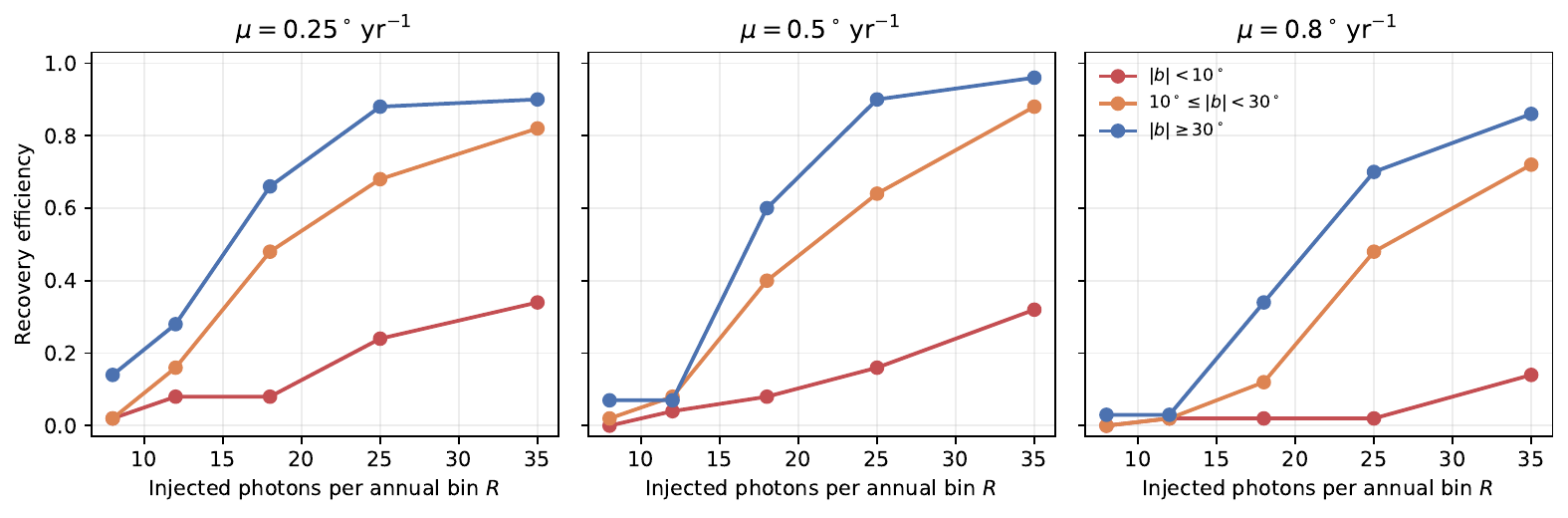}
\vspace{-5mm}
\caption{Injection efficiency in the 45 production cells. Panels show angular speed, and colours show Galactic latitude. Each FL16Y point gives the recovery fraction from 50 trials, balanced across ten representative sky regions. All injections use the incident reference $E^{-2}$ spectrum. \label{fig:efficiency}}
\vspace{-3mm}
\end{figure}

Because the analysis caps overlap, a simulated source may pass through several caps. Each realization is therefore processed in every $18^\circ$ cap crossed by its true path. The same ownership and duplicate-removal rules used for the real and randomized skies are then applied to the reconstructed tracks. An injection is classified as recovered if at least one surviving track matches its known trajectory; additional matching tracks do not change this binary recovery label and are quantified below. The diffuse photon background varies strongly across the sky, particularly near the Galactic plane. We therefore divide the 448 cap centres into three latitude bands, low $ |b|<10^\circ$, intermediate $ 10^\circ\leq |b|<30^\circ$ and high $ |b|\geq30^\circ,$ which contain 78, 146, and 224 centres, respectively. Ten representative sky regions, each corresponding to one $18^\circ$ analysis cap and distributed across Galactic longitude, are selected from each latitude band. For every combination of photon count and angular speed, each region provides five independent source tests.
The completed calibration therefore contains the following number of source tests:
$ 5\ {\rm brightnesses}\times 3\ {\rm speeds}\times 3\ {\rm latitude\ bands}\times 10\ {\rm regions}\times 5\ {\rm repetitions} =2250.$

Let $M$ denote the number of final candidate tracks that match the known trajectory of an injected source. An injection trial is classified as recovered when $M\geq1$, so the efficiency used in the analysis is $\epsilon_{\rm phys}=P(M\geq1).$ This binary efficiency measures whether the source is found at least once; it is not the mean number of candidate tracks produced by that source. At $\mu=0.5\,\mathrm{deg\,yr^{-1}}$, the fractions recovered in the low-, intermediate-, and high-latitude bands are $(8,40,60)\%$ for $R=18$ and $(32,88,96)\%$ for $R=35$. After weighting the three bands by their numbers of sky regions and applying the monotonic efficiency fit, the corresponding full-sky efficiencies are $44.4\%$ and $82.3\%$. The lower efficiency near the Galactic plane reflects the brighter diffuse background, which makes a small moving source harder to distinguish. Faster sources are also generally more difficult to recover because their photons spread farther within each year and their annual positions are more widely separated. The complete recovery-efficiency measurements across all tested photon counts, angular speeds, and Galactic latitude bands are shown in Fig.~\ref{fig:efficiency} for the reference $E^{-2}$ spectrum.

We represent source recovery with a Jeffreys Beta-distribution, thus for $k$ recoveries among $n=50$ source tests, this follows the distribution ${\rm Beta}(k+1/2,n-k+1/2)$. Within the calibrated range, adding more source photons should not reduce the probability of recovery. We therefore require the fitted efficiency to remain constant or increase with $R$, combining neighbouring calibration points when their statistical fluctuations would otherwise reverse this ordering.

\begin{figure}[t]
\centering
\hspace*{-5mm}
\includegraphics[width=0.62\columnwidth]{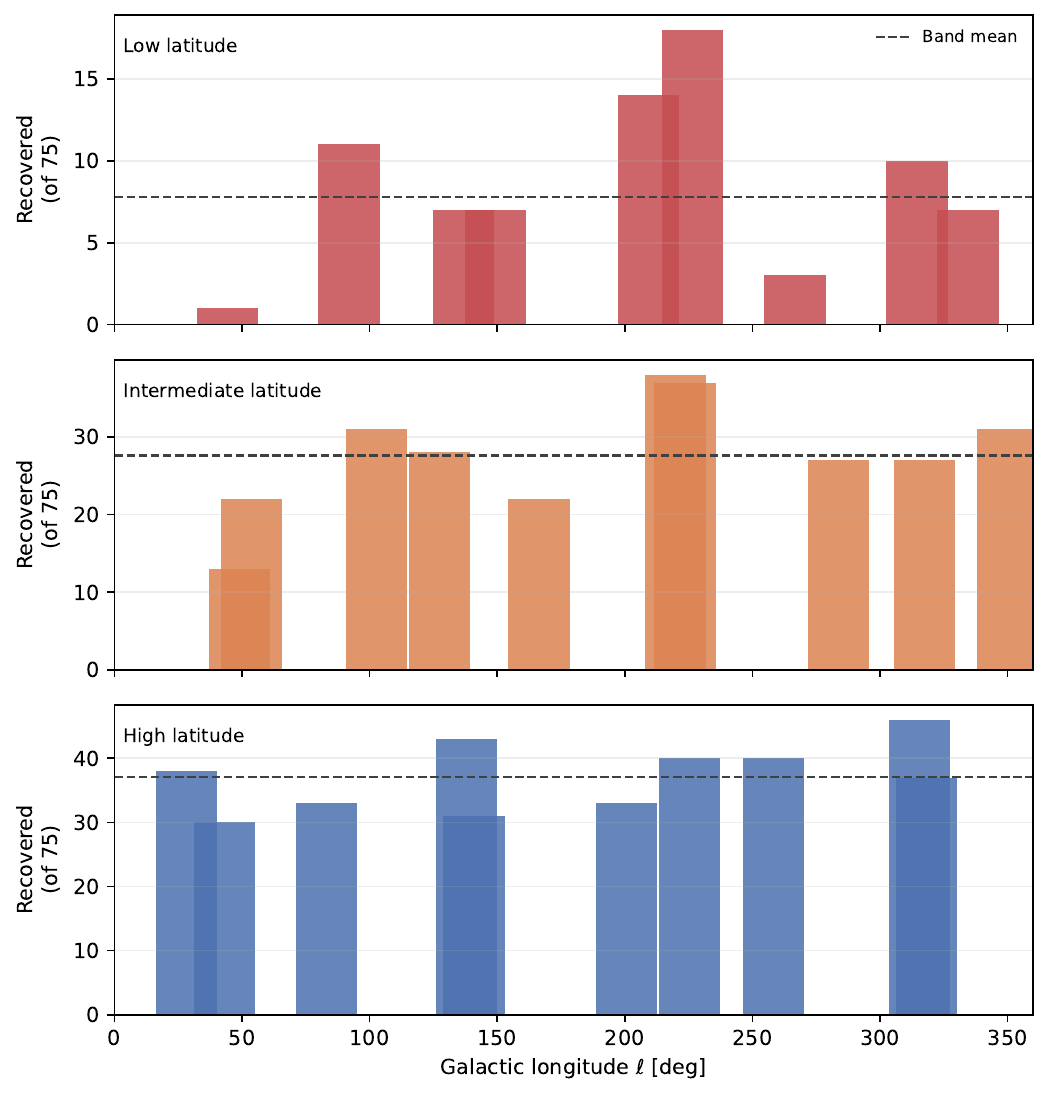}
\vspace{-3mm}
\caption{Recovery grouped by sky region and ordered by Galactic longitude. Each bar shows the number recovered from 75 trials. Panels show the low-, intermediate-, and high-latitude bands in red, orange, and blue, respectively. Dashed lines show the corresponding band means. \label{fig:hosts}}
\vspace{-5mm}
\end{figure}

We performed a separate check with 320 simulated sources to determine why some sources are missed. The search recovers 107 of them. For 126 sources, the photons do not form sufficiently clear annual groups to begin a track. Sixty are rejected because they lie too close to a known catalogue source, 17 do not stand out sufficiently from their local background, and ten form annual groups that cannot be linked into a candidate track satisfying the motion and duplicate-removal requirements. Dense Galactic-plane emission causes 109 of the 126 failures at the first clustering stage. Some positions close to catalogue sources are rejected in every repetition and therefore remain permanent blind spots. Because the main injection calibration follows exactly the same search procedure, these losses reduce the measured efficiency and weaken the resulting limits rather than biasing them.

 The Galactic-plane band contains only 78 of the 448 sky regions, corresponding to 17.4\% of the full-sky weighting. As an extreme test, assuming no sensitivity at all in this band weakens the best exact-count limit from 98.2 to 105 sources, a change of 7.3\%. Conversely, completely correcting the identified clustering loss in dense regions would improve the present limit by less than 10\%. Figure~\ref{fig:hosts} compares the recovery totals for the ten representative regions in each latitude band. Recovery varies with Galactic longitude, but no single region controls the result. Removing any one region changes the best exact-count limit by at most 1.1\%. Resampling the ten complete representative regions within each latitude band gives a separate 90\% interval of 94-102 sources at the best exact-count node.

As a consistency check, we counted every final truth-matched track in the 2,250 production injections. Across this deliberately balanced calibration grid, the 725 injections recovered at least once produced 1,088 matching tracks, corresponding to a mean of $1.50$ tracks per recovered source (95\% interval 1.44-1.56). Our binary efficiency counts each recovered injection only once, thereby underestimating its expected contribution to the track count, making the population translation conservative at the level of the mean signal-track yield. 

\vspace{-5mm}
\subsection{Poisson means and annual exposure}
\label{sec:poisson}
\vspace{-2mm}

A steady astrophysical source does not deliver exactly the same number of detected photons every year. Even at constant intrinsic flux, Poisson counting noise and modest exposure changes make the annual counts fluctuate. To relate the artificial photon counts to a physical source flux, we first calculate the energy-dependent ten-year exposure
\beq
{\cal E}_{10}(E,\Omega) =\int A_{\rm eff}(E,\Omega,t)\,{\rm d}t .
\label{eq:E10}
\eeq
The integral covers all good observing times within the selected ten-year interval. Here $A_{\rm eff}(E,\Omega,t)$ is shorthand for the LAT effective collecting area at energy $E$, evaluated using the orientation of sky direction $\Omega$ relative to the instrument at time $t$. The effective area follows from the public LAT instrument response as a function of photon energy, incidence angle, and event type \cite{Fermi-LAT:2012fsm,Bruel:2018lac}.

 For the reference incident spectrum $dN/dE\propto E^{-2}$, the spectrum-averaged exposure is 
 \beq
 \overline{\cal E}_{10}(\Omega)= \frac{ \int_{1\,{\rm GeV}}^{100\,{\rm GeV}} {\cal E}_{10}(E,\Omega)E^{-2}\,{\rm d}E} {\int_{1\,{\rm GeV}}^{100\,{\rm GeV}} E^{-2}\,{\rm d}E}.
 \eeq
 A source at sky position $\Omega$ with integrated 1-100 GeV photon flux $F_\gamma$ therefore produces an expected ten-year count
 \beq
 \Lambda(F_\gamma,\Omega) =F_\gamma\overline{\cal E}_{10}(\Omega).
 \eeq
 The flux conversion used in the analysis defines the mean annual count as
 \beq
 \lambda(F_\gamma,\Omega) =\frac{\Lambda(F_\gamma,\Omega)}{N_{\rm yr}}, \qquad N_{\rm yr}=10,
 \eeq
where \(N_{\rm yr}=10\) is the number of annual bins. At the median survey exposure, each choice of artificial photon count $ R=(8,12,18,25,35)$ corresponds to fluxes of
$(2.07,~3.11,~4.66,~6.48,~9.07)\times10^{-10} \,\mathrm{ph\,cm^{-2}\,s^{-1}}$, respectively. 
These are convenient labels for the calibration grid rather than five populations with identical physical flux, because the LAT exposure varies across the sky.

The injections described above measure the probability of recovery when exactly $R$ photons are added in every year. A real steady source is different, even if its average brightness remains constant, the number of photons detected in any particular year fluctuates. We therefore use the exact-count results to calculate the recovery efficiency for a source with a specified mean annual count.
Let $T$ be the total number of photons detected over the ten annual bins. Approximating the mean count per bin by $\lambda$, we average the exact-count efficiency over a Poisson distribution with mean $10\lambda$ 
\beq
\epsilon_{\rm phys}(\lambda,\mu,b)= \sum_{T=0}^{\infty} \mathrm{Pois}(T\mid10\lambda)\, \epsilon_{\rm exact}(T/10,\mu,b).
\label{eq:poisson}
\eeq
Here $\epsilon_{\rm exact}(R,\mu,b)$ is the recovery efficiency for an injection with $R$ photons per year, angular speed $\mu$, and Galactic latitude band $b$. Between the five tested values of $R$, we compute the efficiency by interpolating in the logarithm of the photon count. At fixed photon count and Galactic latitude band, we interpolate the recovery efficiency linearly in angular speed between the adjacent tested values. 

The injection analysis measures the recovery efficiency only for sources containing exactly $R=8,12,18,25,35$ photons in each annual bin. We interpolate the exact-count efficiency between these measured points. For the reported physical limits, we conservatively assign zero efficiency when the source's mean annual count satisfies $\lambda<8$ or $\lambda>35$, rather than extrapolating the measured trend. This restriction applies to the source's expected mean count. For a source with $8\leq\lambda\leq35$, Eq.~(\ref{eq:poisson}) averages over fluctuations in the total number of photons detected during the decade. It does not explicitly retain their allocation among individual years; that approximation is tested independently below. The upper cutoff is therefore a boundary of the available calibration, not a claim that brighter sources are physically undetectable. Within the Poisson average, a realization whose decade-averaged count $T/10$ falls below eight is assigned zero efficiency. If $T/10$ exceeds 35 for a source whose mean remains within the calibrated range, the central calculation retains the fitted efficiency at the upper boundary. An alternative treatment assigns zero efficiency to this upper tail to show the uncertainty caused by the finite calibration range. Because the same physical flux produces different expected counts in different parts of the sky, the calculation is performed separately using the exposure of each of the 448 sky regions.

Equation~(\ref{eq:poisson}) predicts the recovery efficiency from the total number of photons detected over the decade. It does not explicitly track how unevenly those photons are distributed among the ten years, which could matter because a valid track must be detected in several annual bins. We therefore tested the approximation using 120 additional simulated sources whose annual photon counts fluctuate independently, with their mean values following the observed year-to-year event-count pattern in the corresponding representative sky region. This pattern varies by 6.16\% rms and serves as a conservative proxy for exposure variation because it can also contain genuine changes in the sky photon rate. These sources span twelve combinations of mean brightness, angular speed, and Galactic latitude, with ten sources in each combination. Before examining the validation results, we used the earlier 2,250-source calibration to predict a recovery probability for each new source. For validation source $i$, with mean annual count $\lambda_i$, angular speed $\mu_i$, and latitude band $b_i$, this probability is 
\beq
p_i= \sum_{T=0}^{\infty} \mathrm{Pois}(T\mid10\lambda_i)\, \epsilon_{\rm exact}(T/10,\mu_i,b_i).
\label{eq:p}
\eeq
The Poisson factor gives the probability of detecting $T$ photons over the ten-year interval, while the measured efficiency gives the probability of recovering a source with the corresponding average count $T/10$. Thus the prediction uses only the injection calibration and the properties of the validation source; it is not fitted to the validation result.

As an example, for a source in the Galactic-plane band with $\lambda=25$ photons per year and $\mu=0.25\,\mathrm{deg\,yr^{-1}}$, Eq.~(\ref{eq:p}) gives $p_i=0.235$. Ten such sources therefore contribute 2.35 expected recoveries. Repeating the same calculation for all twelve tested combinations of brightness, speed, and latitude gives 
\beq
N_{\rm rec}^{\rm exp} =\sum_{i=1}^{120}p_i \simeq53.4.
\eeq
Because each source is either recovered or missed, the expected statistical spread is 
\beq
\sigma_{\rm rec} =\left[\sum_{i=1}^{120}p_i(1-p_i)\right]^{1/2} =4.8. 
\eeq
 The complete search for simulated sources recovered 50 of the 120 validation sources. The 50 observed recoveries are $0.70$ standard deviations below the predicted 53.4, and thus consistent with the calibration.
The resulting flux limit shown in the Letter reaches 123 sources at $F_\gamma=6.5\times10^{-10}~\mathrm{ph~cm^{-2}~s^{-1}}$ and $\mu=0.25~\mathrm{deg~yr^{-1}}$, with a 90\% uncertainty interval of 111-136 sources. Each curve holds the apparent angular speed fixed and is restricted to the calibrated mean-count range of 8-35 photons per year. Section~5 instead averages over the range of angular speeds produced by a physical population with different distances and transverse velocities.

For an isotropic population with common photon flux $F_\gamma$ and apparent angular speed $\mu$, we average the recovery probability over the 448 sky regions, to obtain the physical efficiency
\beq
\overline\epsilon_{\rm phys}(F_\gamma,\mu)=
\frac{1}{N_{\rm sky}}\sum_{i=0}^{N_{\rm sky}-1}
\epsilon_{\rm phys} \left[
\lambda(F_\gamma,\Omega_i),\mu,b_i
\right],
\eeq
and the full-sky number of physical sources 
\beq
N_{90}(F_\gamma,\mu)=
\frac{s_{90}}{\overline\epsilon_{\rm phys}(F_\gamma,\mu)}.
\eeq
Here $\Omega_i$ denotes the centre of sky region $i$, and $\epsilon_{\rm phys}=P(M\geq1)$ is a binary recovery probability. As shown in Section 4, counting each recovered source only once provides a conservative translation of the track-count limit. The calibrated mean-count restriction is applied separately in each region before taking the average. This limit assumes that source positions are uniformly distributed in solid angle.

\vspace{-2mm}
\section{5.~Effective volume and population limits}
\vspace{-2mm}

The injection calibration gives the probability of recovering a source with a specified apparent brightness, angular speed, and sky position. To constrain a physical population, we average this probability over the possible positions and velocities of its members. The result is an effective volume, an ordinary spatial volume weighted by the probability that a source at each position would be detected. A perfectly efficient survey with a sharp distance boundary would recover the corresponding geometric volume. In practice, diffuse backgrounds, catalogue masks, photon-count fluctuations, and search-selection losses make the effective volume smaller.
 
For heliocentric distance $r$ over the full observing time $N_{\rm yr}\simeq10$, a source with photon luminosity $L_\gamma$ in the 1-100 GeV band has an expected number of detected photons per analysis year
\beq
\overline R(L_\gamma,r,\Omega)= \frac{L_\gamma\overline{\cal E}_{10}(\Omega)} {4\pi r^2N_{\rm yr}},
\eeq
where $\overline{\cal E}_{10}(\Omega)$ is the spectrum-averaged ten-year LAT exposure in sky direction $\Omega$. Its apparent angular speed is $\mu=v_\perp/r$, with the quantities converted to consistent angular and time units. We define the probability that this source is recovered as
\beq
P_{\rm det}(L_\gamma,r,\Omega,v_\perp) = \epsilon_{\rm phys} \left[ \overline R(L_\gamma,r,\Omega), \frac{v_\perp}{r}, b(\Omega) \right].
\eeq
Here $b(\Omega)$ is the Galactic latitude and $\epsilon_{\rm phys}$ is the source-recovery efficiency obtained from the injection calibration after including Poisson fluctuations in the detected photon counts. Consequently, $P_{\rm det}$ includes the effects of exposure, angular motion, sky-dependent backgrounds, catalogue masking, and every selection stage of the search. The effective volume for photon luminosity $L_\gamma$ is then
 \beq
 V_{\rm eff}(L_\gamma)= \int {\rm d}\Omega\int_0^\infty r^2{\rm d}r \int_0^\infty {\rm d}v_\perp\, p(v_\perp\mid\Omega)\, P_{\rm det}(L_\gamma,r,\Omega,v_\perp). 
 \eeq
 The velocity distribution $p(v_\perp\mid\Omega)$ is normalized to one in every sky direction, so $V_{\rm eff}$ has the units of an ordinary volume. This expression also shows why brightness and motion cannot be treated separately: increasing the distance reduces the expected photon count as $r^{-2}$ and the angular speed as $r^{-1}$.

The calibration covers expected annual counts $\overline R_{\min}=8$ to $\overline R_{\max}=35$ as well as angular speeds $\mu_{\min}=0.25$ to $\mu_{\max}=0.8\,\mathrm{deg\,yr^{-1}}$. For a fixed sky direction and transverse speed, these ranges define the nearest and farthest distances over which the response has been calibrated:
\beq
r_{\rm near}=\max\left[ \sqrt{\frac{L_\gamma\overline{\cal E}_{10}(\Omega)}{4\pi N_{\rm yr}\overline R_{\max}}}, \frac{v_\perp}{\mu_{\max}} \right],
\quad\quad\quad 
r_{\rm far}=\min\left[ \sqrt{\frac{L_\gamma\overline{\cal E}_{10}(\Omega)} {4\pi N_{\rm yr}\overline R_{\min}}}, \frac{v_\perp}{\mu_{\min}} \right]. 
\eeq
We do not assume that every source has one fixed transverse speed. Instead, we average over a standard model for local Galactic velocities. In the Galactic rest frame, the two transverse velocity components are taken to be Gaussian, with dispersion $\sigma_v=238/\sqrt{2}=168\,\mathrm{km\,s^{-1}}$. We then include the motion of the Sun, using a local-standard-of-rest speed of $232\,\mathrm{km\,s^{-1}}$ and Solar peculiar velocity $(U,V,W)=(11.1,12.24,7.25)\,\mathrm{km\,s^{-1}}$ \cite{Schoenrich:2009bx,McMillan:2016jtx}. Because the component of the Solar motion across a line of sight depends on sky direction, the resulting transverse-speed distribution is
\beq 
p(v_\perp\mid\Omega)= \frac{v_\perp}{\sigma_v^2} \exp\left[-\frac{v_\perp^2+\nu(\Omega)^2}{2\sigma_v^2}\right] I_0\left(\frac{v_\perp\nu(\Omega)}{\sigma_v^2}\right),
\eeq
where $\nu(\Omega)$ is the Solar velocity projected across the line of sight and $I_0$ is the modified Bessel function. This distribution accounts for the fact that the same population has different apparent transverse motions in different parts of the sky. The velocity model affects only the physical interpretation and is not used to select candidate tracks. 

\begin{figure}[t]
\centering
\includegraphics[width=0.65\columnwidth]{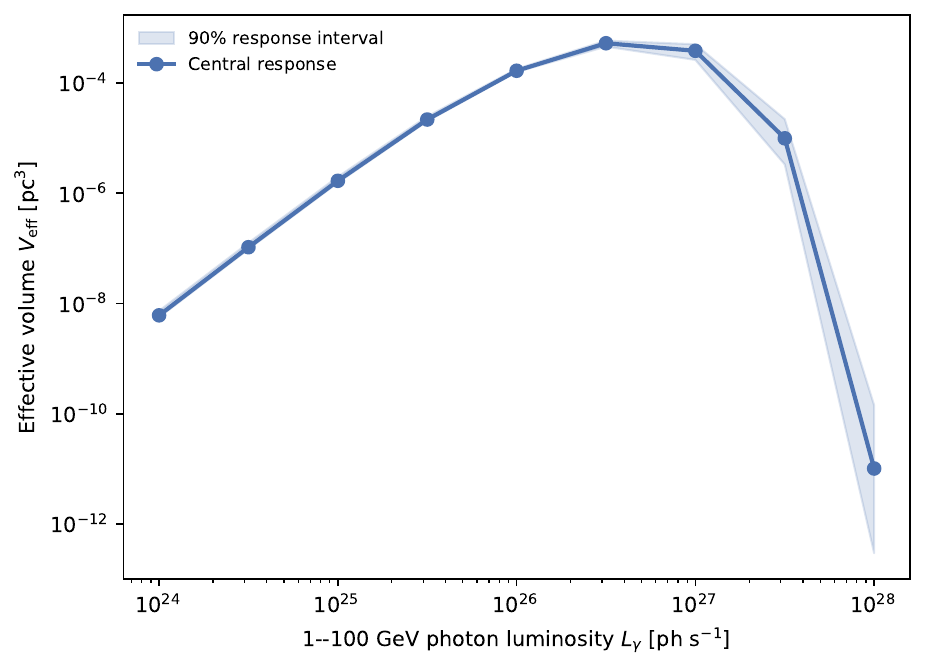}
\caption{Heliocentric effective volume for steady $E^{-2}$ emitters after Poisson convolution. The band propagates uncertainties in the recovery efficiency, exposure, and velocity model. The high-luminosity decline marks finite calibration support and must not be interpreted as decreasing physical detectability.}
\label{fig:veff}
\end{figure}

In Figure \ref{fig:veff} we show the effective volume as a function of luminosity for steady $E^{-2}$ emitting sources.
The effective volume initially grows with luminosity because a brighter source can be detected at greater distances. At high luminosity it declines because the allowed distance range encounters both boundaries of the calibration: nearby sources exceed the 35-photon count ceiling, while more distant sources move more slowly than $0.25\,\mathrm{deg\,yr^{-1}}$. This turnover is therefore a boundary of the calibrated search, not evidence that intrinsically brighter sources are harder to detect. The true effective volume outside this range may be larger, but we do not extrapolate it. We propagate the statistical and astrophysical uncertainties using 20,000 joint realizations for the source-number limits and 2,000 realizations for the effective-volume and abundance calculations.

The randomized-sky contribution is obtained by resampling the 60 complete campaigns with replacement and recalculating $s_{90}$. For each photon-count, angular-speed, and Galactic-latitude cell containing $k$ recoveries in $n$ trials, the injection efficiency is drawn independently from the Jeffreys posterior ${\rm Beta}(k+1/2,n-k+1/2)$. In every realization, the efficiencies are then fitted as a nondecreasing function of photon count using weighted isotonic regression. As a separate robustness check on correlations between injections sharing the same sky region, we generated 100,000 bootstrap samples by resampling the ten complete representative regions with replacement within each latitude band and refitting the monotonic efficiency curves. We report this test of region-to-region variation separately. The uncertainty ranges quoted for the main results are based on the finite number of source-injection trials and do not include an additional contribution from this check.

 The distribution of LAT exposures is propagated by resampling the 448 complete sky regions with replacement within each Galactic-latitude band. For the velocity model, we use eight stratified realizations in which the Galactic halo speed scale is sampled uniformly over $218-258\,\mathrm{km\,s^{-1}}$ and the local-standard-of-rest speed over $220-244\,\mathrm{km\,s^{-1}}$. The Solar peculiar velocity is held fixed at $(U,V,W)=(11.1,~12.2,~7.3)\,\mathrm{km\,s^{-1}}$. For the abundance limits, the local dark matter density is sampled uniformly over $0.3-0.5\,\mathrm{GeV\,cm^{-3}}$, with $0.4\,\mathrm{GeV\,cm^{-3}}$ used for the central result.

We recompute the source-number limits, effective volume, and abundance limits for the corresponding joint realizations. The quoted uncertainty intervals are the central 90\% intervals, defined by the 5th and 95th percentiles; the median is also recorded, while the quoted central curve is evaluated using the central input values. Uncertainty in $s_{90}$ enters the source-number and abundance limits, but not the effective volume itself.

 For the abundance interpretation, we take the local dark matter density to be $\rho_{\rm DM}=0.4\,\mathrm{GeV\,cm^{-3}}$ and vary it over $0.3-0.5\,\mathrm{GeV\,cm^{-3}}$ \cite{Read:2014qva,McMillan:2016jtx}. Neither this density nor the velocity model is used in the search itself.
Let $B_i$ be the number of tracks found in randomized campaign $i$, and let $S(s)$ be a Poisson-distributed number of recovered signal tracks with mean $s$. We define $s_{90}$ as the signal mean for which there is only a 10\% probability of obtaining no more tracks than the observed count, $N_{\rm obs}=577$. Including the finite-campaign plus-one correction, it satisfies 
\beq
\frac{1+\sum_{i=1}^{60} {\rm Pr} \left[S(s_{90})\leq N_{\rm obs}-B_i\right]} {60+1}=0.1.
\eeq
Solving this equation gives $s_{90}=80.8$. Thus a population producing a larger mean number of recovered signal tracks would have less than a 10\% chance of yielding as few total tracks as observed. Resampling the 60 randomized campaigns gives a 90\% uncertainty range of $73.2-86.3$. Importantly, $s_{90}$ constrains the number of signal tracks recovered by the search, rather than directly constraining the number of physical sources.

The 90\% confidence-level upper limit on the full-sky number of physical sources $ N_{90}(R,\mu)$ with exact injected count $R$ per annual bin and angular speed $\mu$ is obtained by correcting the upper limit on recovered signal tracks, $s_{90}$, for the recovery efficiency $\epsilon(R,\mu)$
\beq
N_{90}(R,\mu)=\frac{s_{90}}{\epsilon(R,\mu)}.
\eeq
The most sensitive tested point has $\epsilon=0.823$, for sources producing exactly 35 detected photons in every annual bin and moving at $0.5\,\mathrm{deg\,yr^{-1}}$. This gives $N_{90}=98.2$ sources, with a combined 90\% uncertainty range of $88.8-107$. This is a calibration benchmark rather than a robust flux limit, exposure differs across the sky, and a steady source naturally produces different photon counts from year to year. Both effects are included in $V_{\rm eff}$.

\begin{figure}[t]
\centering
\includegraphics[width=0.6\columnwidth]{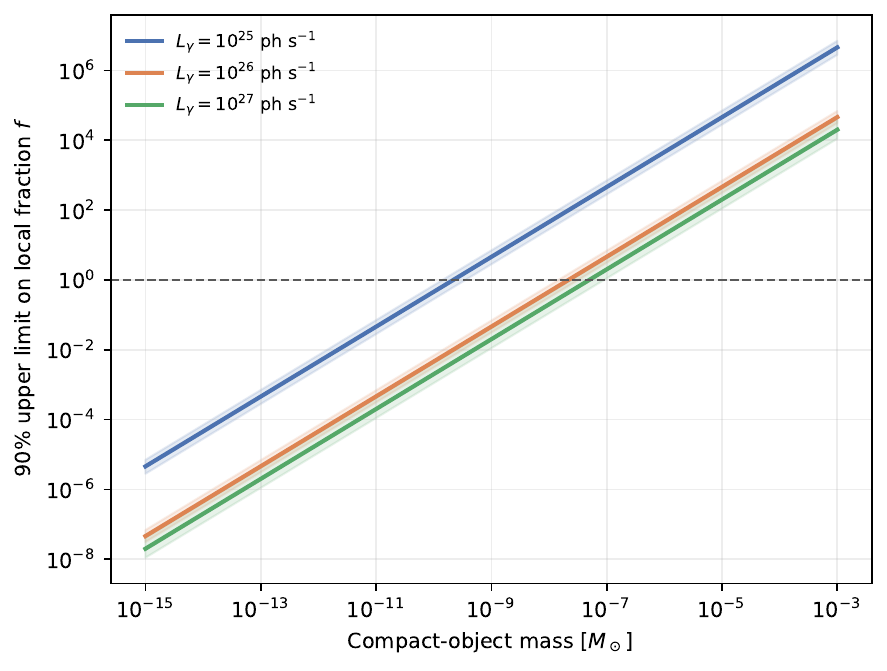}
\caption{Generic 90\% upper limits on the fraction of local dark matter in a monochromatic population of steady compact emitters. Curves correspond to fixed 1--100 GeV photon luminosities, with shaded bands showing the propagated 90\% uncertainty intervals. The dashed horizontal line marks $f=1$.
\label{fig:generic}}
\end{figure}

 For a population with approximately uniform local number density $n$, the expected number of detected sources is $nV_{\rm eff}$. If identical objects of mass $M$ and photon luminosity $L_\gamma$ make up a fraction $f$ of the local dark matter, their number density is $n=f\rho_{\rm DM}/M$. Requiring the expected detected count not to exceed $s_{90}$ gives
 \beq
 N_{\rm det}=\frac{f\rho_{\rm DM}}{M}V_{\rm eff}(L_\gamma),\qquad f_{90}=\frac{s_{90}M} {\rho_{\rm DM}V_{\rm eff}(L_\gamma)}.
 \eeq
 For reference, the central effective volumes at $L_\gamma=10^{25},10^{26},10^{27}\,\mathrm{ph\,s^{-1}}$ are $1.68\times10^{-6}$, $1.67\times10^{-4}$, and $3.83\times10^{-4}\,\mathrm{pc^3}$, respectively. The resulting generic mass-luminosity limits are shown in Fig.~\ref{fig:generic}.

\vspace{-3mm}\section{6.~Illustrative dressed-PBH benchmark}
\vspace{-3mm}

The released computational kernel can be applied to any source within the calibrated signal family: a steady, point-like emitter whose incident photon spectrum is well approximated by the reference $E^{-2}$ distribution. As an illustrative example, we consider PBHs surrounded by dense halos of particle dark matter. The PBH's gravity binds the halo, and annihilations between the dark matter particles produce gamma rays. 
For self-conjugate dark matter with velocity-independent annihilation, the volume-integrated annihilation $J$ factor and photon luminosity are
\beq 
J=\int \rho_\chi^2( x)d^3x, 
\qquad 
\qquad 
L_\gamma=\frac{\langle\sigma v\rangle}{2m_\chi^2}N_\gamma J, 
\eeq
where $N_\gamma$ is the photon yield per annihilation in the 1-100 GeV band. This factorization separates the particle model from the halo structure and allows one to substitute a different mass, cross section, final state, or density profile.
The illustrative benchmark uses a monochromatic PBH population and self-conjugate WIMP dark matter with mass $m_\chi=1$ TeV, 100\% annihilation to $b\bar b$, and $\langle\sigma v\rangle=3\times10^{-26}\,\mathrm{cm^3\,s^{-1}}$ with $N_\gamma\approx45.8$ photons per annihilation \cite{Cirelli:2010xx}.

Dark matter falling into the PBH potential before matter-radiation equality is modeled as a steep spike \cite{Eroshenko:2016yve,Adamek:2019gns}
\beq 
r_{\rm ta}(t_{\rm eq})&=(2GM_{\rm PBH}t_{\rm eq}^2)^{1/3},
\quad\quad 
\rho_{\rm sp}(r)&=\frac{\rho_{\rm eq}}{2} \left[\frac{r_{\rm ta}(t_{\rm eq})}{r}\right]^{9/4}, 
\eeq
with $t_{\rm eq}=2.40\times10^{12}\,\mathrm{s}$ and
$\rho_{\rm eq}=2.10\times10^{-19}\,\mathrm{g\,cm^{-3}}$. Dark matter annihilation leads to a central density plateau at
\beq 
\rho_{\rm ann}=\frac{m_\chi}{\langle\sigma v\rangle t_0} =7.65\times10^{10}\,\mathrm{GeV\,cm^{-3}}
\eeq
for $t_0=13.8$ Gyr. This plateau prevents the formal $r^{-9/4}$ cusp from producing an extreme annihilation rate, wherever the spike would exceed $\rho_{\rm ann}$, annihilations deplete it to the plateau density. The halo model therefore uses $\rho_\chi=\min(\rho_{\rm sp},\rho_{\rm ann})$ between $4r_S$ and $r_{\rm ta}$. It assumes the density profile survives Galactic tides and stellar encounters and has no luminosity gain from later outer accretion \cite{Adamek:2019gns}.

To define the model uncertainty bracket shown in Fig.~4  of the main text, we parameterise the spike normalization relative to the profile above as $A_\rho$. The central model uses $A_\rho=1$, a hard annihilation plateau   $\rho_\chi=\min(\rho_{\rm sp},\rho_{\rm ann})$, and $r_{\rm out}=r_{\rm ta}$. The lower luminosity endpoint uses $A_\rho=1$, a smooth plateau $\rho_\chi=(\rho_{\rm sp}^{-1}+\rho_{\rm ann}^{-1})^{-1}$, and severe tidal truncation at $r_{\rm out}=0.01r_{\rm ta}$. The upper luminosity endpoint uses $A_\rho=1.53$, the hard plateau, and $r_{\rm out}=r_{\rm ta}$. 
The shaded region is a conservative outer envelope that combines this halo-model bracket with the 90\% analysis ranges in $s_{90}$ and $V_{\rm eff}$ and with $\rho_{\rm DM}=0.3$--$0.5\,\mathrm{GeV\,cm^{-3}}$. The shaded region is thus indicative, rather than a strict statistical 90\% confidence interval.

For this profile the plateau radius scales as $M_{\rm PBH}^{1/3}$ and its volume scales with the PBH mass. The volume-integrated annihilation factor and photon luminosity are approximately linear in $M_{\rm PBH}$
\beq 
J&\simeq 7.9\times10^{56} \left(\frac{M_{\rm PBH}}{10^{-9}M_{\odot}}\right) \mathrm{GeV^2\,cm^{-3}}, \quad {\rm and} \quad
L_\gamma&\simeq 5.5\times10^{26} \left(\frac{M_{\rm PBH}}{10^{-9}M_{\odot}}\right) \mathrm{ph\,s^{-1}}. 
\eeq
We consider the mass interval $10^{-10}\leq M_{\rm PBH}/M_\odot\leq10^{-8}$. This benchmark assumes that WIMP thermal motions do not substantially modify the profile near the annihilation core, which dominates $J$. At sufficiently low PBH masses, these motions alter the inner profile, with the transition depending on the WIMP kinetic-decoupling history \cite{Adamek:2019gns}. We do not extrapolate the benchmark below the stated mass interval.
The strongest central bound is $f_{\rm PBH}<8.5\times10^{-3}$ at $M_{\rm PBH}\simeq1.9\times10^{-10}M_{\odot}$.
The weakening in the limits at high mass occurs because the increasing luminosity moves detectable objects toward the count and low angular speed boundaries, not because intrinsically brighter PBHs become harder to detect.
This example shows how the search can constrain one physically motivated PBH model. Its predicted luminosities fall within the range for which the search sensitivity has been calibrated. We calculate the physical gamma-ray output from annihilation to $b\bar b$, but evaluate its detectability using injections calibrated with the reference $E^{-2}$ photon distribution. Thus it is not a general exclusion of PBHs and a dedicated calibration using the full $b\bar b$ spectrum would be required for a spectrum-specific limit.

\vspace{-3mm}
\section{7.~Illustrative quasi-steady Hawking benchmark}
\vspace{-3mm}

An evaporating PBH becomes hotter and more luminous as it loses mass \cite{Hawking:1974rv}. Searches for the final burst thus probe a different signal from the steady moving sources considered in the previous section. We restrict this translation to PBHs whose remaining lifetime is long enough that their 1-100 GeV output changes only modestly over the ten-year data interval. We describe the emission using the analytic, lifetime-integrated Hawking/QCD photon spectrum adopted by the Fermi-LAT PBH search \cite{Fermi-LAT:2018pfs}. We then pass its decade-averaged 1-100 GeV luminosity through the same steady-source response as an illustrative comparison.

For a nonrotating PBH emitting Standard-Model particles, we use the standard Hawking mass-temperature relation and remaining-lifetime approximation \cite{Hawking:1974rv,Fermi-LAT:2018pfs}:
  \beq
M(T)=\frac{1.06\times10^{13}\,\mathrm{g\,GeV}}{T},
\qquad {\rm and } \qquad
\tau(T)=K_\tau T^{-3},
\eeq
where $K_\tau\approx 4.8\times10^{11}\,\mathrm{s\,GeV^3}$ corresponds to a $10^{10}\,\mathrm{g}$ black hole having a remaining lifetime of $400\,\mathrm{s}$. Here $T$ and $\tau(T)$ are evaluated at the beginning of the observing interval.

Integrating the analytic broken-power-law approximation to the lifetime-integrated Hawking/QCD photon spectrum \cite{Fermi-LAT:2018pfs} over this energy band gives
\beq
{\cal N}_{1-100}(T)
={\cal A} \times
\begin{cases}
\dfrac{1}{2}\left(E_-^{-2}-E_+^{-2}\right),
& T\leq E_-,
\\[2mm]
2T^{-3/2}\left(E_-^{-1/2}-T^{-1/2}\right)
+\dfrac{1}{2}\left(T^{-2}-E_+^{-2}\right),
& E_-<T<E_+,
\\[2mm]
2T^{-3/2}\left(E_-^{-1/2}-E_+^{-1/2}\right),
& T\geq E_+ .
\end{cases}
\label{eq:remaining-hawking-photons}
\eeq
We take $E_-=1\,\mathrm{GeV}$ and $E_+=100\,\mathrm{GeV}$ and ${\cal A}=0.45\times9\times10^{35}$ is the numerical normalization adopted for the central benchmark following \cite{Fermi-LAT:2018pfs}.
The model variation, leading to the shaded region in Fig.~5  of the main text, corresponds to deviating from this central normalization by taking ${\cal A}=(0.32$-$0.58)\times9\times10^{35}$. The shaded region combines this model normalization with the 90\% analysis ranges in $s_{90}$ and $V_{\rm eff}$. Since the normalization range is a model bracket rather than a probability distribution, the resulting envelope is not itself a 90\% confidence interval.

During an observing interval $\Delta t<\tau(T)$, the temperature rises to
\beq
T_{\rm end}
=T\left[\frac{\tau(T)}{\tau(T)-\Delta t}\right]^{1/3}.
\eeq
The mean 1-100 GeV photon luminosity over that interval is therefore
\beq
\overline L_\gamma(T)=
\frac{{\cal N}_{1-100}(T)-{\cal N}_{1-100}(T_{\rm end})}
{\Delta t},
\eeq
where $\Delta t$ is expressed in seconds when $\overline L_\gamma$ is quoted in $\mathrm{ph\,s^{-1}}$.
We require a remaining lifetime $\tau\geq100$ yr, ten times the data interval, and verify that the instantaneous band luminosity changes by less than 25\% during the decade. For illustration, we evaluate temperatures $T\simeq2.6-5.3$ GeV, corresponding to masses $M\simeq(2-4)\times10^{12}$ g and remaining lifetimes of roughly 100-830 yr. The lower-temperature endpoint is set by the lower luminosity boundary of the calibrated effective-volume response, rather than by the quasi-steady requirement.

For PBHs with current temperature $T$ and an approximately uniform local number density $n(T)$, the expected number of recovered sources is 
\beq
N_{\rm det}(T)=n(T)\, V_{\mathrm{eff}}[\overline L_\gamma(T)].
\eeq
Here $V_{\mathrm{eff}}$ accounts for the distances, sky positions, transverse velocities, exposure, and search efficiency over which a PBH with decade-averaged luminosity $\overline L_\gamma(T)$ could be detected. Requiring the expected number of detections not to exceed the 90\% upper limit $s_{90}$ gives
\beq
n_{90}(T)= \frac{s_{90}} {V_{\mathrm{eff}}[\overline L_\gamma(T)]}.
\eeq
This is a limit on the present-day local number density of PBHs in the specified quasi-steady temperature, mass, and remaining-lifetime interval, rather than on the total density of PBHs at all evolutionary stages. Across the relevant temperature interval, $V_{\mathrm{eff}}[\overline L_\gamma(T)]$ ranges from $1.40\times10^{-8}$ to $5.13\times10^{-7}\,\mathrm{pc^3}$, giving number density limits from $5.78\times10^9$ down to $1.57\times10^8\,\mathrm{pc^{-3}}$ as the PBH temperature increases. 

We derive a bound on the number density, but this applies only to PBHs that are in this short, slowly changing stage today. We can also constrain the local rate at which PBHs enter and pass through this stage. If that rate $\dot n$ is approximately steady, the expected number of detections is obtained by adding the effective volume available at each temperature, weighted by the time a PBH spends there:
\beq
N_{\rm det}&=\dot n\int_{T_{\min}}^{T_{\max}} V_{\mathrm{eff}}\left[\overline L_\gamma(T)\right] \left|\frac{d\tau}{dT}\right|dT, 
\eeq
 where $\tau$ is expressed in years in this integral, so
\beq
\left|\frac{d\tau}{dT}\right|
=3\left(\frac{K_\tau}{1\,\mathrm{yr}}\right)T^{-4}
\simeq 4.5\times10^{4}\,\mathrm{yr\,GeV^3}T^{-4}.
\eeq
Assuming a steady local population of PBHs passing through the validated 100-830 yr remaining-lifetime interval according to the adopted Hawking emission model, this corresponds to a 90\% upper limit $\dot n_{90}\simeq1.36\times10^6\,\mathrm{pc^{-3}\,yr^{-1}}$ on the local evaporation rate. 

Finally, as discussed in the main text, the current analysis should be viewed as an illustrative luminosity-based translation rather than a spectrum-specific Hawking-radiation exclusion: the physical Hawking spectrum enters through its integrated 1-100 GeV luminosity, which is evaluated using the common reference source response.

\end{document}